\documentclass[useAMS,usenatbib]{mn2e}
\usepackage{epsfig,amsmath,natbib}
\usepackage{color}
\usepackage[bookmarks,bookmarksopen,colorlinks,linkcolor={blue},citecolor={red},urlcolor={green}]{hyperref}

\def\gsim{\lower.5ex\hbox{\gtsima}}
\def\lsim{\lower.5ex\hbox{\ltsima}}
\newcommand{\cmfast}{\textsc{\tt 21cmFAST}}
\newcommand{\medea}{\textsc{\tt MEDEA2}}
\newcommand{\camb}{\textsc{\tt CAMB}}

\def\be{\begin{equation}}
\def\ee{\end{equation}}

\def\gtsima{$\; \buildrel > \over \sim \;$}
\def\ltsima{$\; \buildrel < \over \sim \;$}
\def\prosima{$\; \buildrel \propto \over \sim \;$}
\def\gsim{\lower.5ex\hbox{\gtsima}}
\def\lsim{\lower.5ex\hbox{\ltsima}}
\def\simgt{\lower.5ex\hbox{\gtsima}}
\def\simlt{\lower.5ex\hbox{\ltsima}}
\def\simpr{\lower.5ex\hbox{\prosima}}
\def\la{\lsim}
\def\ga{\gsim}

\newcommand{\de}{{\rm d}}
\newenvironment{packed_enum}{
\begin{enumerate}
  \setlength{\itemsep}{0pt}
  \setlength{\parskip}{0pt}
  \setlength{\parsep}{0pt}
}{\end{enumerate}}

\title[The nature of DM from the global high-z HI 21 cm signal]{The nature of dark matter from the global high redshift HI 21 cm signal}         
\author[Vald\'{e}s, Evoli, Mesinger, Ferrara \& Yoshida]
{M. Vald\'{e}s$^{1,3}$\thanks{E-mail: marcos.valdes@sns.it}, C. Evoli$^{2,1}$, A. Mesinger$^1$, A. Ferrara$^{1,3}$ \& N. Yoshida$^{4,3}$\\
$^1$ Scuola Normale Superiore, Piazza dei Cavalieri 7, 56126 Pisa, Italy\\
$^2$ II. Institut f\"{u}r Theoretische Physik, Universit\"{a}t Hamburg, Luruper Chaussee 149, 22761 Hamburg, Germany\\
$^3$ Kavli IPMU, University of Tokyo, 5-1-5 Kashiwanoha, Kashiwa, Chiba 277-8568, Japan\\
$^4$ Department of Physics, University of Tokyo, 3-7-3 Hongo, Bunkyo-ku, Tokyo 113-0033, Japan}

\begin{document}
\maketitle
\label{firstpage}

\begin{abstract}
We study the imprint of dark matter (DM) annihilation on the global 21~cm signal from the Dark Ages to Cosmic Reionization.  Motivated by recent observations, we focus on three DM candidates: (i) a 10 GeV Bino-like neutralino (ii) a 200 GeV Wino and (iii) a 1 TeV heavier particle annihilating into leptons.
For each DM candidate we assume two values for the thermally averaged annihilation cross section $\langle \sigma v \rangle$, the standard thermal value $\langle \sigma v \rangle_{th} = 3 \times 10^{-26} \mbox{cm}^3 \, \mbox{s}^{-1}$ and the maximum value allowed by WMAP7 data, $\langle \sigma v \rangle_{max}$.
 We include the enhancement of DM annihilations due to collapsed structures, detailed estimates of energy deposition into the intergalactic medium (IGM),
 as well realistic prescriptions for astrophysical sources of UV and X-ray radiation.
In these models, the additional heat input from DM annihilation suppresses the mean 21cm brightness temperature offset by $\delta T_b \sim$ few--100 mK.
In particular, the very deep $\delta T_b \sim -150$~mK absorption feature at $\sim$ $20 \lsim z \lsim 25$ predicted by popular models of the first galaxies is considerably reduced or totally erased by some of the considered DM candidates. 
Such an enhancement in IGM heating could come from either DM annihilations or a stronger-than-expected astrophysical component (i.e. abundant early X-ray sources).  However, we find that the two signatures are not degenerate, since the DM heating is dominated by halos several orders of magnitude smaller than those hosting galaxies, whose fractional abundance evolves more slowly resulting in a smaller gradient: $d\delta T_{b}/d\nu \lsim 4$ ${\rm mK}/{\rm MHz}$ in the range $\nu \sim 60-80$ ${\rm MHz}$. The detection of such signals by future radio telescopes would be clear evidence of DM energy injection at high-redshifts.
 \end{abstract}



\section{Introduction}

In the framework of the successful $\Lambda$CDM cosmology theory only $\sim$ 5 \% of the current energy content of the Universe is made of visible matter, while the rest is divided into two {\it dark} components, 73 \% in the form of the so called Dark Energy, with the remaining 22 \% accounted for by Dark Matter (DM) which only interacts with baryons through gravity \citep{Komatsu:2009}. The existence of DM seems to be indirectly confirmed by a large set of observations (even though it is conceivable that in the future a new theory of gravity could do without DM), however its nature remains unknown. The accuracy of $\Lambda$CDM in explaining the evolution of the Universe is one of the major successes of modern cosmology; however the question of what makes up over 95~\% of the Universe is still open.

A huge effort has been made in the past decades to detect DM directly and indirectly. Currently, the best we can do is to place constraints on the properties (e.g. mass, annihilating cross section, decay rate) of some of the proposed DM candidates, thanks mostly to recent observations. We will focus here on Weakly Interacting Massive Particles (WIMPs) that are among the most popular DM candidates since their existence and properties are predicted by several extensions of the Standard Model of particle physics. 

WIMPs have a small but non negligible interaction cross-section with ordinary matter and therefore in principle they can be directly detected via elastic collision with the nuclei of terrestrial targets. Several recent experiments such as DAMA/LIBRA, DAMA/Nal, CDMS-II, EDELWEISS-II, CoGeNT, XENON100 have attempted such detection (see e.g. \citealt{Bernabei:2004, Bernabei:2008, Bernabei:2010, Aalseth:2011a, Aalseth:2011b, Aprile:2011, Ahmed:2010}; Armengaud et al. 2010). Promising results came from CoGeNT, which reported about a hundred events exceeding the expected background, possibly originated from the nuclear recoil by scattering from DM particles of mass $M_{\chi} \lsim 10$ GeV. A light WIMP with similar mass is also favored by the recent DAMA/LIBRA annual modulation signal (\citealt{Bernabei:2010, Aalseth:2011a, Aalseth:2011b}).

Astrophysical observations in the X-ray and Gamma radiation bands have also been used to indirectly detect a DM generated signal. A lot of excitement followed the recent detection by the PAMELA, ATIC, FERMI-LAT and HESS experiments of an excess in the electron/positron cosmic ray energy spectrum, which could be explained by annihilating DM with mass of order of 1 TeV \citep[see e.g.][]{Cirelli:2008, Liu:2009, Berg:2009, Hooper:2009, Chen:2009, Abdo:2010a}. The results appear to be far from conclusive due to the difficulties in properly modeling the cosmic ray energy spectrum and because other astrophysical sources such as pulsars could be responsible for the signal \citep[see e.g.][]{diBernardo:2011, Profumo:2008}.

Another promising observational window for the indirect search of DM will be available in the near future when next generation radio interferometers such as the Low frequency Array (LOFAR\footnote{\url{http://www.lofar.org/}}), the 21 Centimeter Array (21CMA\footnote{\url{http://21cma.bao.ac.cn/}}), the Murchison Widefield Array (MWA\footnote{\url{http://www.mwatelescope.org/}}) and the Square Kilometer Array (SKA\footnote{\url{http://www.skatelescope.org/}}). These large interferometers will be able to map at arcminute scales the redshifted 21~cm line corresponding to the hyperfine triplet-singlet line transition of the ground level neutral hydrogen (HI) at $z>6$, i.e. during the Epoch of Reionization (EoR) and possibly well into the so called Cosmic Dark Ages of the Universe \citep[see e.g.][]{Peterson:2005, Bowman:2006, Kassim:2004, Wyithe:2005}. 

The global sky-averaged 21~cm signal could potentially be measured as a function of frequency even by single dipoles: this is the main scientific aim of current instruments such as the Experiment to Detect the Reionization Step (EDGES), while more advanced ones such as the Large-aperture Experiment to Detect the Dark Ages (LEDA\footnote{\url{http://www.cfa.harvard.edu/LEDA/}}) and the Dark Ages Radio Experiment (DARE\footnote{\url{http://lunar.colorado.edu/dare/mission.html}}), a lunar orbiting dipole experiment, are planned for construction (see e.g. \citealt{Bowman:2010, Greenhill:2012, Burns:2012}).

Any release of energy from DM - for instance, as is the case considered in this work, through annihilations - would be in part absorbed by the intergalactic medium (IGM) as heat and ionization. If this occurred before the first astrophysical sources ($z\gsim30$), it would produce a deviation from the theoretically well established thermal and ionization history \citep[see e.g.][]{Chen:2004, Mapelli:2006, Cirelli:2009} and could alter significantly the HI 21 cm signal \citep[see e.g.][]{Shchekinov:2007, Furlanetto:2006, Valdes:2007, Natarajan:2009, Galli:2011, Finkbeiner:2012}. If the DM energy injection occurred after the first astrophysical sources ($z\lsim30$), the deviation could be stronger and easier to observe, but would have to be disentangled from astrophysical uncertainties concerning the first galaxies.
There are enormous observational challenges in detecting the 21 cm signal at $z \gsim 30$, and such observations may be decades away from our current technological level. For this reason, we focus on the DM signal at lower redshift, after stars and galaxies started forming, and look for strategies to separate the DM signal from that of astrophysical sources.

We investigate the effects produced on the HI 21 cm signal by three DM candidates: (i) a 10 GeV Bino-like neutralino (ii) a 200 GeV Wino and (iii) a 1 TeV heavier particle annihilating into leptons. 
These candidates have been recently proposed to explain the aforementioned indirect/direct hints of DM detection. We allow the annihilation cross-section to be in a range which is compatible with CMB observations. For the first time we compute the 21 cm DM signal at $z \lsim 30$ in a self consistent scenario that includes a realistic prescription for the formation of astrophysical sources of UV and X-ray radiation.

The details of the energy deposition into the IGM from DM annihilations are an essential ingredient to compute the HI 21 cm signal. In \cite{Evoli:2012} we have studied the cascade produced by the products of DM annihilations for the three considered DM candidates: this allows us to discuss the impact of DM annihilations in a physically consistent way and to assess the observability of the chosen DM candidates with HI 21~cm observations in the near future.

The rest of the paper is organized as follows. In Sec. 2 we briefly introduce the DM candidates which we study; in Sec. 3 we explain in detail the method that we follow to compute the 21 cm DM signal. In Sec. 4 we present the results of our calculations, and in Sec. 5 we discuss the results and draw our conclusions.


\section{Dark Matter models}\label{sec:dmmodels}

In this Section we briefly introduce our DM models of choice. For a more detailed discussion of such DM particle candidates we refer the reader to \cite{Evoli:2011}, were they are described extensively. We select three sample cases which have been recently investigated in connection with hints of DM signals in either direct and indirect search experiments. These sample cases are also representative of three different WIMP mass regimes, ranging from fairly light models to multi-TeV DM, and of three different kinds of annihilation channels.

{\bf Wino:} We consider a pure Wino within the Minimal Supersymmetric extension to the Standard Model (MSSM). The recent interest in this model has been stimulated, besides its peculiar signatures at the LHC \citep{Bertone:2012}, by the claim \citep{Grajek:2009, Kane:2009} that a Wino with mass of about 200 GeV can explain the rise detected by PAMELA in the positron fraction. 

{\bf Bino:} We consider the $b\bar{b}$ case with mass of 10 GeV to model a strong coupling with quarks as suggested by recent results of direct detection experiments (e.g. CDMS, DAMA).

{\bf Leptophilic:} Again motivated by the PAMELA positron excess, and possibly in connection with the local all-electron (namely electrons plus positrons) flux measured by Fermi and HESS, several analyses have considered the possibility of very heavy dark matter WIMPs, with masses up to several TeV and very large pair annihilation cross section \citep[see e.g.][]{Cirelli:2009b, Bergstrom:2009}. The results of such studies are that, to account for the electron/positron component without violating the antiproton bounds, dark matter needs to be {\it leptophilic}, i.e. the final products of the annihilation being dominantly leptons, most likely a combination of $e^{+}e^{-}$ and $\mu^{+}\mu^{-}$. It has been recently pointed out \citep{Ciafaloni:2011} that for very heavy WIMPs the radiative emission of soft electroweak gauge bosons is crucial to model the high-energy spectrum. 
\begin{table*}
\caption{The DM models described in Sec.~\ref{sec:dmmodels}.}
\begin{center}
\begin{tabular}{|c|c|c|c|c|c|}
\hline
DM model & mass [GeV] & $\langle \sigma v \rangle$ [cm$^3$/s] & $\epsilon_0$ [eV/s] & $\delta \tau_e$ & Line style \\
\hline
W$^{+}$W$^{-}$ & 200 & $\langle \sigma v \rangle_{th}=3.0 \times 10^{-26}$        & $5.35 \times 10^{-25}$ & $1.53 \times 10^{-3}$ & blue solid \\ 
W$^{+}$W$^{-}$ & 200 & $\langle \sigma v \rangle_{max}=1.2 \times 10^{-24}$   & $2.14 \times 10^{-23}$ & $6.09 \times 10^{-2}$& blue dashed \\
$b\bar{b}$           & 10   & $\langle \sigma v \rangle_{th}=3.0 \times 10^{-26}$       & $1.07 \times 10^{-23}$ & $1.80 \times 10^{-2}$& red solid \\ 
$b\bar{b}$           & 10   & $\langle \sigma v \rangle_{max}=1.0 \times 10^{-25}$  & $3.57 \times 10^{-23}$ & $5.76 \times 10^{-2}$& red dashed \\
$\mu^{+}\mu^{-}$ & 1000 & $\langle \sigma v \rangle_{th}=3.0 \times 10^{-26}$    & $1.07 \times 10^{-25}$ & $1.42 \times 10^{-4}$& green solid \\ 
$\mu^{+}\mu^{-}$ & 1000 & $\langle \sigma v \rangle_{max}=1.4 \times 10^{-23}$ & $4.99 \times 10^{-23}$ & $6.18 \times 10^{-2}$& green dashed \\
\hline
\end{tabular}
\end{center}
\label{tab:edm}
\end{table*}
\section{Method}

We first constrain the properties of our DM candidates from CMB data. This allows us to associate to each particle a range of allowed values for the annihilation cross section $\langle \sigma v \rangle$. 
We then use the fractional energy depositions from \medea\ \citep{Evoli:2012} and the galactic radiation fields from \cmfast\ \citep{Mesinger:2011} in order to compute the thermal and ionization evolution of the IGM and the associated global 21cm signal. We describe each of these steps in turn.

\subsection{DM heating}

Including the formation of substructures at redshift $\lsim$~50 naturally enhances the effects of DM annihilations: while they do not dramatically impact the global reionization history~\citep{Cirelli:2009} the increased energy injection could alter predictions of the observable 21 cm signal by heating the IGM.
The total energy release by DM annihilations per unit volume is given by (see e.g. \citealt{Cirelli:2009, Chluba:2010}):
\begin{equation}\label{eq:chluba}
\frac{\de E_{\chi}}{\de t} = 2 M_{\chi} c^{2} \langle \sigma v \rangle n_{\chi}^{2} (1+B(z)),
\end{equation}
where $\langle \sigma v \rangle$ is the thermally averaged annihilation cross section, $M_{\chi}$ is the DM particle mass and $n_{\chi} = n_{0,\chi}(1+z)^{3}$ is the number density of DM particles and anti-particles with present day value:
\begin{equation}
n_{0,\chi} = 1.2 \times 10^{-8} {\rm cm^{-3}} \left(\frac{\Omega_{\chi}h^{2}}{0.11}\right) \left( \frac{M_{\chi}c^{2}}{100 \, \rm GeV} \right)^{-1}.
\end{equation}
The term $B(z)$ defines the effective averaged DM density resulting from structure formation and can be written as:
\begin{equation}\label{eq:bz}
B(z) = \frac{\Delta_{\rm vir}(z)}{3 \rho_{c} \Omega_{\rm M}} \int_{M_{min}}^{\infty} dM M \frac{dn}{dM}(z,M) F(M,z)
\end{equation}
where $\de n/\de M$ is the halo mass function obtained adopting the Press-Schechter formalism \citep{Press:1974} and $F(M,z)$ is the concentration function depending on the distribution of DM inside halos, a function that peaks typically for sub-halo masses $M_{\rm sh} \sim 1$ M$_{\odot}$ at $z=20$ (see Fig. \ref{fig:concentration}; we refer the reader to \citealt{Cirelli:2009} for details)\footnote{Although extrapolating the halo mass function down to such small masses is highly uncertain, we note that at the scales and redshifts of interest, the Press-Schechter and Sheth-Thormen cumulative mass functions \citep{Press:1974, Sheth:1999, Jenkins:2001} agree to $\sim 10\%$.}. We use here a prescription for the effect of substructures and consider a minimum mass for primordial proto-haloes of $M_{\rm min} = 10^{-6}$ M$_{\odot}$ for the 10 GeV Bino, and  $M_{\rm min} = 10^{-9}$ M$_{\odot}$ for the other two more massive candidates. This choice is motivated by recent calculations of the exponential cutoff mass in the power spectrum as a function of neutralino mass, resulting from free-streaming of the DM particles after the kinetic decoupling (see e.g.\citealt{Bringmann:2009, Aarssen:2012}).

Recent works \citep[e.g.][]{Slatyer:2009, Galli:2009} have investigated the possible role of Sommerfeld enhancements due to which $\langle \sigma v \rangle$ can be considerably boosted and becomes a function of redshift. We choose not to make assumptions on the physical processes that could boost $\langle \sigma v \rangle$ and treat it as a parameter constrained by CMB observations.

We covered until now the energy {\it production} by DM annihilations. The most relevant quantity for our purposes is however the energy {\it deposition} into the IGM.  Only a small fraction of the energy released by DM annihilations is finally deposited into the IGM in the form of heat, excitations and ionizations either of hydrogen and helium. The absorbed fraction depends on the DM candidate - in the form of the initial energy spectrum from the annihilation event - and on the environment where the annihilation takes place, specifically on the ionized fraction of the ambient gas and on the energy density of CMB photons, important for the Inverse Compton (IC) scatterings.

To model this, we use the Monte Carlo scheme \medea\ (for details see \citealt{Evoli:2012}). Through \medea\ we are able to calculate what fraction of the energy released by a single annihilation event goes into: (i) H and He ionizations (ii) excitations (iii) heating. A set of handy fitting formulae that take into account the dependence on $z$ and $x_{\rm e}$ are then given for the respective quantities $f_{ion}(x_{e},z), \, f_{a}(x_{e},z), \, f_{h}(x_{e},z)$.

\subsection{CMB constraints}

%
\begin{figure}
\centering
\includegraphics[width=0.45\textwidth]{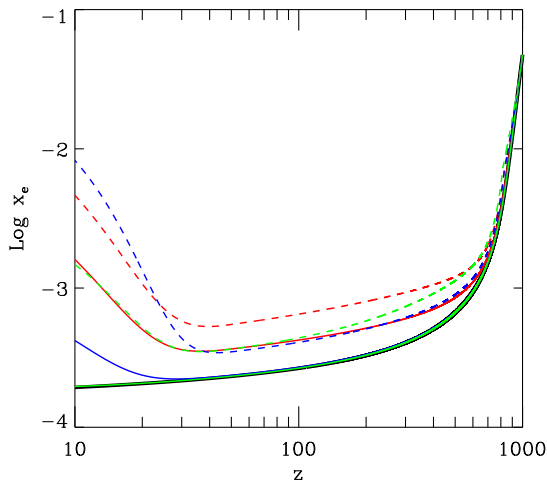}\\
\caption{\small IGM ionization fraction as a function of redshift. Black solid line is the case without annihilating DM, for the color scheme of the DM models see Tab.~\ref{tab:edm}.}\label{fig:xe}
\end{figure}
\begin{figure}
\centering
\includegraphics[width=0.45\textwidth]{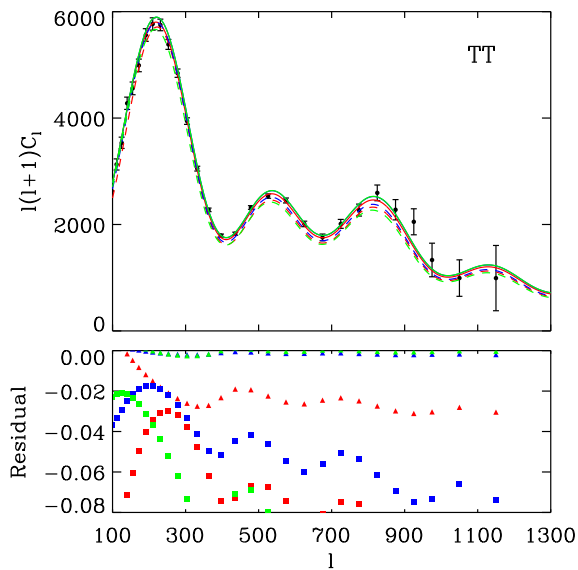}\\
\caption{\small (Top panel) TT CMB power-spectrum for the DM models in Tab.~\ref{tab:edm}. Black solid line is the case without annihilating DM. (Bottom) Residual of the CMB power-spectrum with respect to the case without DM annihilations. The DM models are presented in red (10 GeV); blue (200 GeV); green (1 TeV). The triangles and squares are used for $\langle \sigma v \rangle_{th}$ and $\langle \sigma v \rangle_{max}$ respectively.}\label{fig:cls}
\end{figure}
%

We compute the modifications induced by DM annihilations on the CMB power spectra and on the integrated Thomson optical depth to verify that our models are consistent with WMAP7 observations and to put upper limits on $\langle \sigma v \rangle$ for the considered DM candidates. To do so we modify the public code \camb\ \citep{Lewis:2011} and add a DM term to the evolution equations of the IGM kinetic temperature and ionized fraction (Fig.1).
%
%
 
With this modified version of \camb\ code we can calculate the effects of DM annihilations on the temperature (TT), polarization (EE) and temperature-polarization (TE) CMB spectra. The TT spectrum allows us to put the most sensitive constraints of DM properties (see Fig.2). Our results are obtained assuming that the cosmological parameters have best-fitting values as indicated by the 7-yr WMAP data \citep{Komatsu:2011}. For each DM candidate we increase the value of the thermally averaged annihilation cross section from the standard value $\langle \sigma v \rangle_{th} = 3\times 10^{-26} \rm cm^{3} s^{-1}$ up to a value $\langle \sigma v \rangle_{max}$ for which the computed TT CMB spectrum exceeds by more than $3 \sigma$ from the best-fit the WMAP7 data. The upper limits $\langle \sigma v \rangle_{max}$ for each of the considered DM candidates are given in Tab.~\ref{tab:edm} along with the color and line-style later used in the plots.
CMB observations by Planck will allow to put more stringent constraints on the properties of the DM candidates in the near future \citep{Planck:2006, Galli:2009}.

We then calculate the contribution of DM to the Thomson optical depth, $\delta \tau_{e}$, by integration:
\begin{equation}
\delta \tau_{e} = \int_{6}^{10^{3}} c \sigma_{T} \delta x_{e}(z) N_{b}(z) \left| \frac{\de t}{\de z} \right|  dz
\end{equation}
where $\delta x_{e}$ is the difference in the ionized fraction between the standard scenario and the case in which DM annihilates. We assume here that the Universe is fully ionized at redshift $z<6$: this contributes by a factor $\tau_{\rm e}\approx 0.04$ to the WMAP measured total optical depth $ \tau_e = 0.088 \pm 0.015$ \citep{Komatsu:2011, Shull:2012}. 
Note that DM annihilations are only imprinted on the CMB as an additional source of ionization.  Therefore the above procedure is effectively the same as choosing an upper limit for the extra contribution to $\tau_e$ from DM annihilations.  From Tab.~\ref{tab:edm} we see that this is $\delta \tau_{e} \approx 0.06$, making the total optical depth (in the absence of any additional astrophysical sources at $z>6$): $\tau_e \approx 0.1$.  This limit is conservatively low, roughly corresponding to the 1-$\sigma$ upper limit from WMAP7 obtained from the TE cross-correlation \citep{Komatsu:2011}.

Notice that the heating/ionization contribution of the two heavier DM candidates assuming $\langle \sigma v \rangle_{max}$ increases proportionally by a larger factor over the case with $\langle \sigma v \rangle_{th}$ than what found for the 10 GeV Bino when assuming $\langle \sigma v \rangle_{max}$ rather than $\langle \sigma v \rangle_{th}$. This is due to our assumption of a smaller minimum halo mass $M_{\rm min} = 10^{-9}$ M$_{\odot}$ that enhances strongly the effects of substructures and to which the CMB constraints are, on the other hand, less sensitive.

\subsection{21 cm signal}

One of the observable quantities most likely to carry a trace of the effects of DM annihilations is the redshifted 21~cm line associated with the hyperfine transition between the triplet and the singlet levels of the neutral hydrogen ground state.  This signal is most commonly expressed in terms of the differential brightness temperature between a neutral hydrogen patch and the CMB (neglecting redshift-space distortions):
\begin{align}
\label{eq:delT}
\nonumber \delta T_{b} = &\frac{T_S - T_{\rm CMB}}{1+z} (1 - e^{-\tau}) \approx \\
\nonumber &27 x_{\rm HI} (1+\delta) \left(1 - \frac{T_{\rm CMB}}{T_S} \right) \\
&\times \left( \frac{1+z}{10} \frac{0.15}{\Omega_{\rm M} h^2}\right)^{1/2} \left( \frac{\Omega_b h^2}{0.023} \right) {\rm mK},
\end{align}

\noindent where $x_{\rm HI}$ is the neutral fraction of the gas, $\delta$ the overdensity, and $T_S$ is the spin temperature which is set by the number densities of hydrogen atoms in the singlet ($n_0$) and triplet ($n_1$) ground hyperfine levels, $n_{1} /n_{0}=3 \exp (- 0.068$K$/T_{S})$.

It is theoretically well known that in the presence of the CMB alone, $T_S$ reaches thermal equilibrium with $T_{\rm CMB}=2.73 \, (1+z)$~K on a short time-scale, making the HI undetectable in emission or absorption. However, collisions and scattering of Ly$\alpha$ photons $-$ the so-called Wouthuysen-Field process or Ly$\alpha$ pumping (e.g. \citealt{Wouthuysen:1952, Field:1959, Hirata:2006}) can couple $T_S$ to $T_K$.

The spin temperature can be written as (e.g. \citealt{Furlanetto:2006r}):
\begin{equation}
\label{eq:Ts}
{T_S}^{-1} = \frac{ {T_{\rm CMB}}^{-1} + x_{\alpha} {T_{\alpha}}^{-1} + x_c {T_K}^{-1} }{1 + x_{\alpha} + x_c}
\end{equation}
where $T_{\alpha}$ is the color temperature, which is closely coupled to $T_K$ \citep{Field:1959}, and $x_{\alpha}$ and $x_c$ are the two coupling coefficients relative to Ly$\alpha$ scattering and collisions respectively. 
If either collisions or Ly$\alpha$ radiation couple $T_S$ to $T_K$ then the neutral hydrogen will be visible in absorption or emission depending on whether the gas is colder or hotter that the CMB respectively.

For details about the physics associated with the HI 21 cm line and with the determination of $T_S$, the quantity that governs it, we refer the reader to, e.g., \cite{Furlanetto:2006r, Hirata:2006, Valdes:2008}. For our purposes it is important to state here that the physical quantities that determine $T_S$ are: (i) the gas density $n_H$; (ii) the CMB temperature $T_{\rm CMB}$; (iii) the kinetic temperature of the gas, $T_K$; (iv) the ionized fraction $x_e$; and (v) the Ly$\alpha$ background intensity $J_{\alpha}$.

While we know the average evolution of $n_H$ and $T_{\rm CMB}$ we have to determine the others as a function of redshift. The equations that describe the average evolution of the ionized fraction $x_{e}$ and of the kinetic temperature $T_K$ are the following \citep[see e.g. ][]{Chen:2004, Mesinger:2011}:   
\begin{equation}
\label{eq:ion_rateacc}
\frac{dx_e}{dz} = \frac{dt}{dz} \left[ \Gamma_{\rm ion}
  - \alpha_{\rm B} x_e^2 n_b f_{\rm H} \right] ~ ,
\end{equation}
\begin{align}
\label{eq:dTkdzacc}
\nonumber \frac{dT_K}{dz} &= \frac{2T_K}{1+z} +  \frac{2}{3 k_B (1+f_{He}+x_{e})} \frac{dt}{dz} \sum_p \epsilon_p~ ,
\end{align}
where $n_b=n_{b, 0} (1+z)^3$ is the mean baryon number density, $\epsilon_p (z)$ is the heating rate per baryon for each process $p$ in erg s$^{-1}$, $\Gamma_{\rm ion}$ is the ionization rate per baryon, $\alpha_{\rm B}$ is the case-B recombination coefficient, $k_B$ is the Boltzmann constant and $f_{\rm He}$ is the helium fraction by mass. 

The term $\Gamma_{\rm ion}$ includes both the contribution from galaxies and the term that accounts for DM annihilations. Similarly we have that $\epsilon_p (z)=\epsilon_{\rm CMB} (z)+\epsilon_{\rm X}(z)+\epsilon_{\rm DM}(z)$ is the sum of three contributions: (i) $\epsilon_{\rm CMB}(z)$ is Compton heating from CMB photons; (ii) $\epsilon_{\rm X}(z)$ is heating from astrophysical sources, which we take to be dominated by X-rays; (iii) $\epsilon_{\rm DM}(z)$ is DM heating. Notice that when including DM in the terms $\Gamma_{\rm ion}$ and $\epsilon_p (z)$ we use the specific fractional energy depositions $f_h (x_e,z), \, f_{ion} (x_e,z)$ from \cite{Evoli:2012}. 

The last equation needed to compute the 21~cm background is the one describing the evolution of the Ly$\alpha$ background intensity $J_\alpha$:
\begin{equation}
J_{\alpha} = J_{\alpha,R} + J_{\alpha,C} + J_{\alpha,\ast} + J_{\alpha,X} + J_{\alpha,DM} ,
\end{equation}
where the contributions on the RHS correspond to recombinations, collisional excitations by electron impacts, direct stellar emission, X-ray excitation of HI, and DM annihilations (respectively) \citep[see e.g.][]{Madau:1997, Valdes:2007, Mesinger:2011}. The coupling coefficient, $x_\alpha$ is proportional to the  Lyman $\alpha$ background flux, $J_{\alpha}$ (e.g. \citealt{Furlanetto:2006r}).  

In general, if we neglect the energy input from DM, we expect $T_K$ and $T_{S}$ to track $T_{\rm CMB}$ down to $z \sim 300$, when $T_K$ decouples from $T_{\rm CMB}$ and starts decreasing adiabatically as the Universe expands. $T_S$ is then coupled to $T_K$ due to the high gas density and the consequent strong collisional coupling. At $z \sim 70$ $T_S$ gradually couples to $T_{\rm CMB}$ until, at $z \sim 30$ becoming virtually identical to it. It is believed however that at around this redshift the first collapsed luminous sources would ignite. Radiation produced by the first galaxies starts to ionize and heat the gas: on a short timescale, at a redshift $z_{\rm WF} \sim 25-30$, the Ly$\alpha$ pumping effectively couples $T_S$ to $T_K$, and only later, at $z_{\rm heat} \sim 18-22$, heating from galaxies drives $T_K$ to values higher than $T_{\rm CMB}$, making the neutral regions in the IGM visible in emission. Therefore a second absorption feature is expected at $z_{\rm heat} \lsim z \lsim z_{\rm WF}$ (see e.g. \citealt{Furlanetto:2006r, Pritchard:2008}). We can divide the global evolution of $\delta T_b$ in six main phases:
\begin{enumerate}
\item 
$\delta T_b = 0$ for z \gsim 300;
\item 
$\delta T_b < 0$ for $30 \lsim z \lsim 300$: this extended absorption feature has a minimum of $\sim 45$~mK at $z \sim 90$;
\item 
$\delta T_b \sim 0$ for $z_{\rm WF} \lsim z \lsim 30$;
\item 
$\delta T_b < 0$ at $z_{\rm heat} \lsim z \lsim z_{\rm WF}$, a second absorption feature less extended in redshift than the previous one but deeper, with $\delta T_{b,min} \sim -150$~mK;
\item 
$\delta T_b > 0$, i.e. in emission, due to heating by galaxies at $z \lsim z_{\rm heat}$ down to the Epoch of Reionization at $z = z_{\rm EoR} \sim 6-8$;  
\item 
$\delta T_b \sim 0$ for $0 \lsim z \lsim z_{\rm EoR}$, where we still have a signal only from self-shielded systems.
\end{enumerate} 

Introducing the effects of DM annihilations in the standard scenario described above can produce deviations on $\delta T_{b}$: in particular the second absorption feature can be strongly modified by energy release by DM annihilations, at a redshift range $15 \lsim z \lsim 25$ that will be probed by the next generation of radio observatories such as LEDA and SKA.

\subsection{Radiation from astrophysical sources}

\begin{figure*}
{\centerline{\includegraphics[width=17truecm]{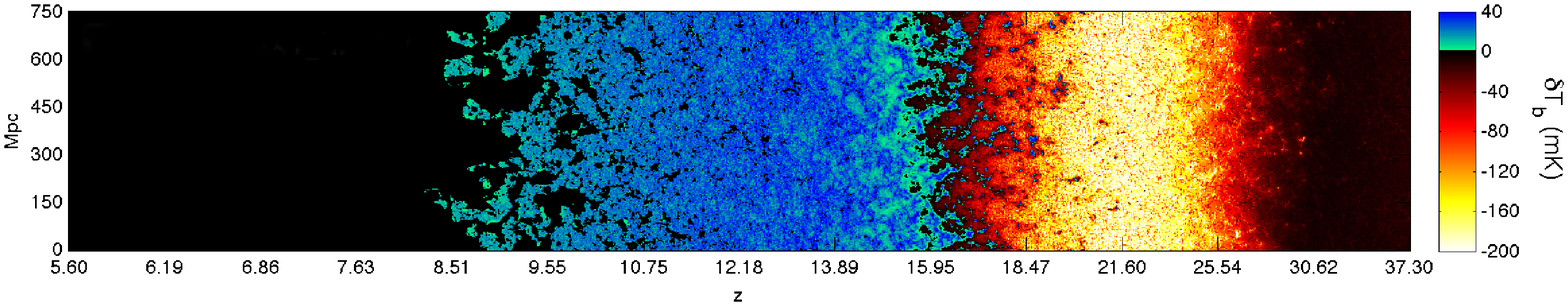}}}
{\centerline{\includegraphics[width=17truecm]{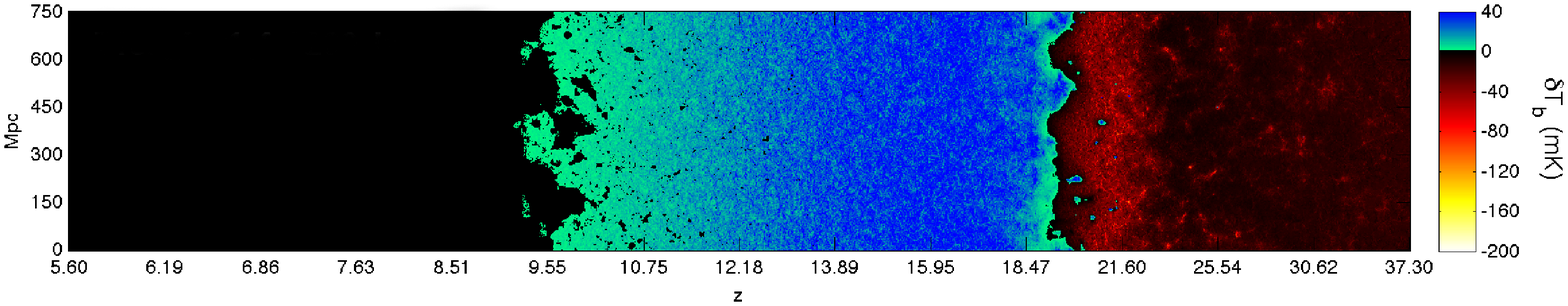}}}
\caption{
Slices through the 21cm brightness temperature maps for two models with different prescriptions for the galaxy properties \citep{Mesinger:2012a}. The simulations are 750 Mpc on a side, with a resolution of $500^3$.  Each slice is 1 cell (1.5 Mpc) thick. The horizontal axis shows evolution along the comoving line-of-sight coordinate. The top panel corresponds to a ``fiducial'' model, in which the X-ray luminosity of primordial galaxies is the same as that observed in nearby starburst galaxies, while the lower panel corresponds to a model in which primordial galaxies are much more efficient in generating X-rays, saturating the soft X-ray background at $z\gsim10$ (see text for details). 
}\label{fig:models}
\end{figure*}

To compute the astrophysical contribution to $\Gamma_{\rm ion}$, $\epsilon_p$, and $J_{\alpha}$, we use the publicly available code, \cmfast\ \footnote{\url{http://homepage.sns.it/mesinger/Sim.html}}.  This code uses a combination of perturbation theory and excursion-set formalism to compute various cosmic fields, and is in good agreement with radiative transfer simulations of reionization \citep[e.g. ][]{Mesinger:2007, Zahn:2011, Mesinger:2011}. The code is fully described in \cite{Mesinger:2011}, to which we refer the interested reader for more details.  Here we briefly note that the code takes into account inhomogeneous X-ray ionization and heating, as well as Ly$\alpha$ pumping from the first UV sources, integrating the evolution of cosmic structures and radiation fields along the light cone.  Although \cmfast\ computes 3D realizations, in this work we only study the global 21cm signal, deferring analysis of the spatial structure to future work.

The simulation boxes are 750 Mpc on a side, with a final resolution of $500^3$. The initial conditions are sampled on a $1500^3$ grid. In Fig.~\ref{fig:models} are presented the 1-cell thick (1.5 Mpc deep) slices through the HI 21~cm brightness temperature maps for two models with different prescriptions for the galaxy properties \citep{Mesinger:2012a}. The top panel corresponds to a ``fiducial''\footnote{Notice that our "fiducial" model does not exactly correspond to the fiducial model in \citealt{Mesinger:2012a} (consistent with the recent measurement of the $0.5 - 8$~keV X-ray luminosity of star forming galaxies from \citealt{Mineo:2012}), but instead corresponds to their T1e4\_fuv1\_fx5\_1keV model (based on an extrapolation of the $2-10$~keV data from \citealt{Gilfanov:2004}). This model represents a more conservative choice since the astrophysical heating is 4 times higher, thus making the DM annihilation signal less apparent.} model, in which the X-ray luminosity of primordial galaxies is the same as that observed in nearby starburst galaxies (e.g. \citealt{Furlanetto:2006a} and references therein). The lower panel corresponds to an "extreme" model in which primordial galaxies, albeit rarer and appearing later, were much more efficient in generating hard X-rays. The later scenario is inspired by recent theoretical (e.g. \citealt{Linden:2010, Mirabel:2011}) and observational \citep{Reichardt:2011, Mesinger:2012} claims. More specifically, these two models have the following characteristics:
\begin{packed_enum}
\item {\it fiducial} (top panel): Galaxies hosting UV and X-ray sources reside in atomically-cooled halos with virial temperatures $T_{\rm vir} > 10^4$ K (corresponding to halo masses of $M_{\rm halo} > 3\times10^7 M_\odot$ at $z\approx20$).  Ly$\alpha$ pumping (i.e. spin temperature coupling)  is dominated by early UV sources, assuming PopII stellar spectra (e.g. \citealt{Barkana:2005}) and a 10\% efficiency of conversion of gas into stars.  The X-ray luminosity of galaxies follows a $h\nu^{-1.5}$ power law shape with a lower limit of $h\nu_0=$ 300 eV (e.g. \citealt{Madau:2004}), and an X-ray efficiency corresponding to $\sim$2 X-ray photons per stellar baryon.  Similar models, inspired by lower-redshift X-ray binary-dominated star-burst galaxies, have been explored in prior analytic studies (e.g. \citealt{Furlanetto:2006a, Pritchard:2007}).\\
\item {\it extreme} (bottom panel): Galaxies hosting UV and X-ray sources reside in more massive halos, similar to ones observed at moderate redshifts (e.g. \citealt{Labbe:2010}) and which are more resilient to feedback effects (e.g. \citealt{Springel:2003, Mesinger:2008, Okamoto:2008}). These galaxies have virial temperatures $T_{\rm vir} > 10^5$ K, corresponding to halo masses of $M_{\rm halo} > 10^9 M_\odot$ at $z\approx20$.  Interestingly, in this case Ly$\alpha$ pumping is dominated by the X-ray excitation of HI, and has a very different spatial signature from the fiducial model.
The X-ray luminosity of galaxies also follows a $h\nu^{-1.5}$ power law shape, but with a more energetic lower limit of $h\nu_0=$ 900 eV (corresponding to heavy obscuration, e.g. \citealt{Lutovinov:2005}), and an X-ray efficiency corresponding to $\sim$4000 X-ray photons per stellar baryon.  This scenario is considered ``extreme'' since the $z\gsim10$ X-ray sources dominate reionization and saturate the unresolved soft X-ray background (\citealt{HM07, Mesinger:2012a}).
\end{packed_enum}
Both scenarios are consistent with the WMAP7 constraints on $\tau_e$ at 2$\sigma$ \citep{Komatsu:2011}.  
Notice that the values of $\langle \sigma v \rangle_{max}$ computed in Section 3.2 represent an upper limit and are not self-consistent with the contribution to $\tau_e$ from the astrophysical sources modelled by \cmfast\ . On the other hand the DM models with the thermal cross-section $\langle \sigma v \rangle_{th}$ are fully consistent since they only contribute negligibly to $\tau_e$.

It is obvious from Fig. \ref{fig:models} that the fundamental epochs in cosmic evolution show very different 21cm signatures in the two models: spin temperature (WF) coupling (black $\rightarrow$ yellow); X-ray heating of the IGM (yellow $\rightarrow$ blue); reionization (blue $\rightarrow$ black).
Here we focus on the global HI 21 cm signal as a way to constrain DM and save spatial signature for future work.


\section{Results}

We present here the results obtained for the three DM models introduced in Sec.~\ref{sec:dmmodels}, each with two values for the annihilation cross $\langle \sigma v \rangle$: the standard thermal value, and the maximum allowed by CMB constraints as described in Sec 3.2. We solve the evolution equations of the kinetic temperature and ionized fraction of the IGM for each of these DM candidates using the fractional energy depositions from the code \medea\, and use a prescription for the formation of collapsed sources of radiation from the publicly available code \cmfast\ , that allows us to include consistently X-ray heating, ionization and Ly$\alpha$ pumping from galaxies. 

\begin{figure}
{\centerline{\includegraphics[width=0.5\textwidth]{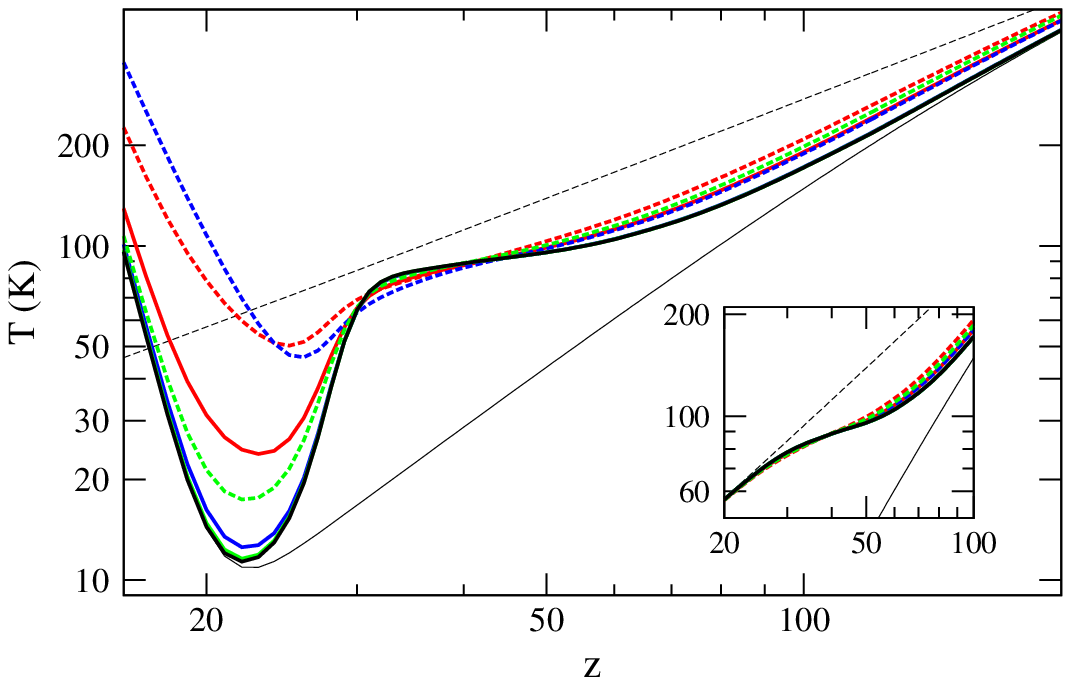}}}
{\centerline{\includegraphics[width=0.5\textwidth]{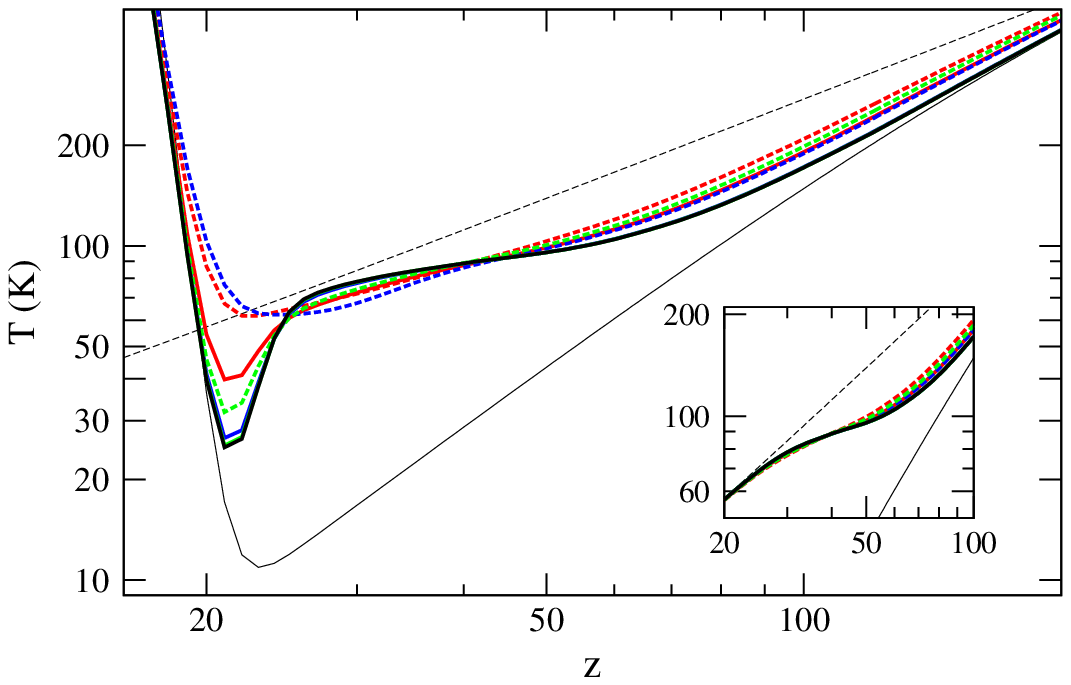}}}
\caption{\small $T_{S}$, $T_k$ (thin black solid line) and $T_{\rm CMB}$ (black short-dashed line) as a function of redshift. The colored $T_S$ curves show the modified behavior due to DM annihilations following the color scheme given in Tab. 1. The solid thick black line is the standard $T_S$ without DM energy input. The top and bottom panels represent the calculations done for the {\it fiducial} and {\it extreme} models respectively. The box on the lower right corner in each panel shows the behavior of $T_S$ for the considered DM candidates without including radiation from luminous sources.}\label{fig:temps}
\label{}
\end{figure}

In Fig.~\ref{fig:temps} we compare the effects produced on $T_{S}$ by our annihilating DM candidates. 
To keep the figure simple we don't plot the modified $T_K$ curves, corresponding to the annihilating DM cases. However the behavior is simple: the kinetic temperature at high redshift increases because of the high density of DM and the corresponding higher chance of annihilations. As the Universe expands and its content is diluted $T_K$ simply settles on a slightly higher adiabat. When the Ly$\alpha$ pumping becomes efficient at $z \lsim z_{\rm WF}$, then $T_S$ perfectly tracks $T_K$, which increases sharply at $z \lsim z_{\rm heat}$.

The additional heating from annihilating DM particles drives $T_S$ closer to $T_{\rm CMB}$ for $40<z<100$. Therefore we can expect a reduction of the absorption feature of the homogeneous 21 cm background in that redshift range. A larger thermally averaged cross-section corresponds to stronger effects on $T_S$ for an annihilating DM candidate of a given mass. At the same time, heavier DM candidates deposit their energy into the IGM less efficiently (see \citealt{Evoli:2012}), therefore given a certain value of $ \langle \sigma v \rangle$ the lower mass DM candidate will generally produce larger deviations on $T_S$. The effects are much more evident at $z \lsim 30$: galaxies start to form and quickly drive $J_{\alpha}$ to values high enough for efficient Ly$\alpha$ pumping. As a consequence $T_S$ decreases sharply tracking $T_K$ at a time when the combined heating from DM annihilations and from galaxies is still not sufficient to heat the gas above $T_{\rm CMB}$. Eventually the same heating sources make $T_K \sim T_S > T_{\rm CMB}$. The different models in the upper and lower panels of Fig. \ref{fig:temps} show different behaviors since in the "extreme" case Ly$\alpha$ coupling is achieved later, and the heating starts earlier, making the region at $16 \lsim z \lsim 30$ in which $T_S < T_{\rm CMB}$ both shallower (in K) and narrower (in redshift).

This is reflected directly on the behavior of $\delta T_b$, shown in the panels of Fig.~\ref{fig:dtb}, for which we give a quantitative analysis case by case. A summary of the results is available in Table 2.

\subsection{10 GeV Bino}

The lightest considered DM candidate, the 10 GeV Bino, produces the largest signature in the "fiducial" model (top panel of Fig. \ref{fig:dtb}), both when assuming a standard thermal cross section $\langle \sigma v \rangle_{th} = 3 \times 10^{-26} \mbox{cm}^3 \, \mbox{s}^{-1}$ (solid red curves in Fig.~\ref{fig:dtb}) and when taking into account the maximum cross section allowed by CMB data, in this case $ \langle \sigma v \rangle_{max} = 1.0 \times 10^{-25} \mbox{cm}^3 \, \mbox{s}^{-1}$ (dashed red curves). In the first case the signal is $\delta T_b \sim 4 - 10$~mK on a redshift range $45 \lsim z \lsim 300$, with a peak of $\sim 10$ mK at $z \sim 100$, and is much more substantial for the second absorption feature at $16 \lsim z \lsim 30$, with a deviation with respect to the standard case in which DM does not annihilate (which we denote hereafter $\delta T_{b,0}$) of $\Delta T_{b,DM} \equiv | \delta T_{b} - \delta T_{b,0} | \sim 100$ mK, a large signal at frequencies $\nu \sim 80$ MHz. The effect is enhanced essentially by a factor two for the higher allowed annihilation cross section and reaches values of $\Delta T_{b,DM} \sim 20$ mK at $45 \lsim z \lsim 300$, while the second absorption feature at lower redshift is essentially erased, with the IGM appearing in even emission already by $z \lsim 25$. These very large signals, both for $ \langle \sigma v \rangle_{max}$ and for $ \langle \sigma v \rangle_{th}$ could be detected by future radio observations (see Section 4.5). In the "extreme" model case (bottom panel) the DM signature before the first astrophysical sources turn on ($z \sim 45-300$) is obviously identical to the "fiducial" model case. The second absorption feature instead changes substantially and is both shallower, with a minimum value of $\delta T_{b} \sim -60$~mK at $z \sim 21$, and narrower in redshift at $z \sim 18 - 25$. This reflects on the DM signal: when assuming $\langle \sigma v \rangle_{th}$ ($\langle \sigma v \rangle_{max}$), $\Delta T_{b,DM} \sim 30 \,\, (45)$~mK at $z \sim 21$.

\subsection{200 GeV Wino}

\begin{figure}
{\centerline{\includegraphics[width=0.5\textwidth]{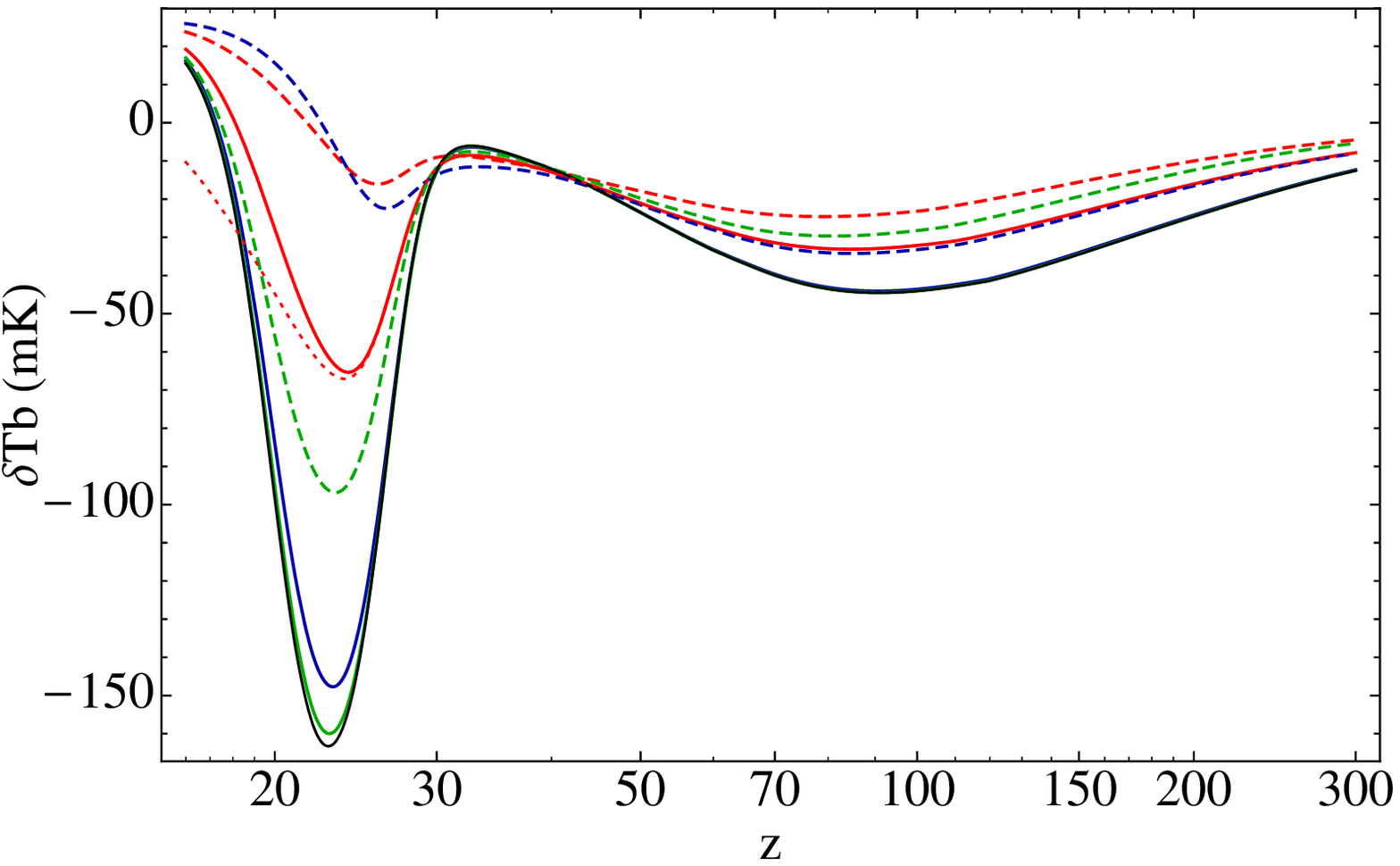}}}
{\centerline{\includegraphics[width=0.5\textwidth]{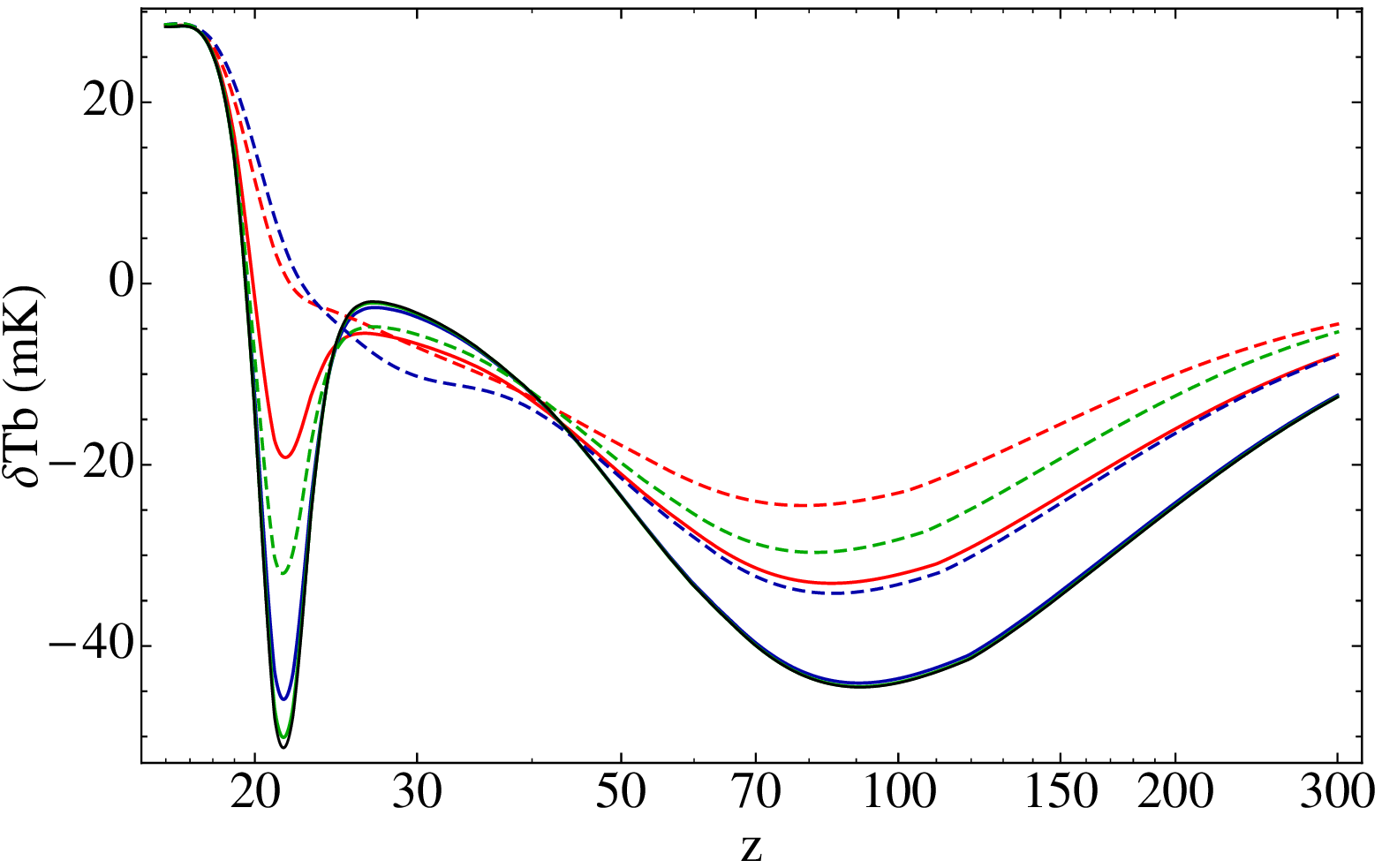}}}
\caption{\small $\delta T_b$ as a function of redshift for all the considered DM models. The standard case with no energy input from DM annihilations is represented with the thick black line. The colored lines follow the scheme given in Tab. 1. The top and bottom panels represent the calculations done for the {\it fiducial} and {\it extreme} models respectively. The red dotted curve in the upper panel corresponds to the thermal Bino model, but ignoring the X-rays from astrophysical sources (i.e. assuming astrophysical sources of X-rays ignited at later times). Notice that some of the curves relative to the DM models are below the black curves at $z=30-40$. This is due to the term $J_{\alpha, DM}$, i.e. the extra Ly$\alpha$ coupling from the DM annihilations.}\label{fig:dtb}
\end{figure}

The case of the 200 GeV Wino, described by the blue curves in Fig.~\ref{fig:dtb} produces less evident deviations: assuming $\langle \sigma v \rangle_{th}$ we can see no effect on the $\delta T_b$ as the solid green line is virtually coincident with the solid black line for both reionization models. Obviously this translates into a constant $\Delta T_{b,DM} = 0$, with no chances of direct detection. On the other hand the case $\langle \sigma v \rangle_{max} = 1.2 \times 10^{-24} \mbox{cm}^3 \, \mbox{s}^{-1}$ shows a $\Delta T_{b,DM}= 2 - 9$~mK for $60 \lsim z \lsim 300$ and a massive deviation at $z \lsim 23$ that practically erases the second absorption feature and drives the signal to emission already at $z \sim 20$. This behavior is present for both the fiducial and extreme reionization models.

\subsection{1 TeV Leptophilic}

Our most massive candidate, the heavy DM particle of rest mass 1 TeV that pair annihilates into leptons, is the one that produces by far the smallest deviations on the HI 21 cm background: for $\langle \sigma v \rangle_{th}$ deviations are negligible while for the maximum allowed cross section, in this case $\langle \sigma v \rangle_{max} = 1.4 \times 10^{-23} \mbox{cm}^3 \, \mbox{s}^{-1}$, they reach $\Delta T_{b,DM} \sim 15$~mK in the range $60 \lsim z \lsim 300$ and $\Delta T_{b,DM} \sim$ 65 (25) mK at $20 \lsim z \lsim 30$ for the fiducial (extreme) reionization models.

As mentioned previously, the fractional increase in the DM signal for the two heavier candidates assuming $\langle \sigma v \rangle_{max}$ rather than $\langle \sigma v \rangle_{th}$ is larger than what found for the 10 GeV Bino, due to our choice for the heavier DM particles of a smaller minimum sub-halo mass $M_{\rm min} = 10^{-9}$ M$_{\odot}$.

\subsection{Isolating the DM annihilation signal}

\subsubsection{Degeneracy with astrophysics}

Although the additional heat input from DM annihilation is easy to isolate in the well-understood epoch of the Dark Ages (before the first astrophysical sources, $z\gsim$ 30--50), the Earth's ionosphere makes observations of such high redshifts very challenging (though there are a couple of non-terrestrial radio telescopes being considered, such as DARE and the Lunar Radio Array\footnote{\url{http://lunar.colorado.edu/lowfreq/}}, LRA).  The lower-redshift signal at $z_{heat} < z < z_{WF}$ is easier to observe; however in this case the DM annihilation heating must be disentangled from uncertainties in astrophysics.  This is evident when comparing the top and bottom panels of Fig. \ref{fig:dtb}.  In the ``extreme'' model in the lower panel the emission from astrophysical sources produces a much shallower absorption feature, and the black solid line could be confused with the red solid line (corresponding to the case of 10 GeV Bino annihilations with $ \langle \sigma v \rangle_{th}$) in the upper panel. Therefore detecting a $\sim -50$~mK global absorption signal at $z \sim 22$ could be an indication of 10 GeV DM annihilations or of a strong X-ray emission by primordial galaxies. Notice however that the X-ray emission can be constrained in several ways, e.g. by tighter constraints on the unresolved soft X-ray background of or by a study of the topology of the HII regions once the future radio interferometers start probing the tomography of the IGM beyond the EoR. Unfortunately, both of these observations are only indirect. The soft X-ray background at $z=0$ only constrains the high-energy component of the source's spectrum ($\sim$ 20--40 keV at $z\sim20$), which does not interact with the IGM (e.g. \citealt{McQuinn:2012}).  Likewise, the EoR at $z\sim10$ could probe a different source population than was present at $z\sim20$ (e.g. \citealt{RO:2004}). Nonetheless, if an upper limit is given to the efficiency of the X-ray emission by primordial galaxies a strong reduction of the depth of the absorption feature would be a clear indication of DM annihilations.

\subsubsection{DM signal gradient}

\begin{figure}
\centering
\includegraphics[width=0.41\textwidth]{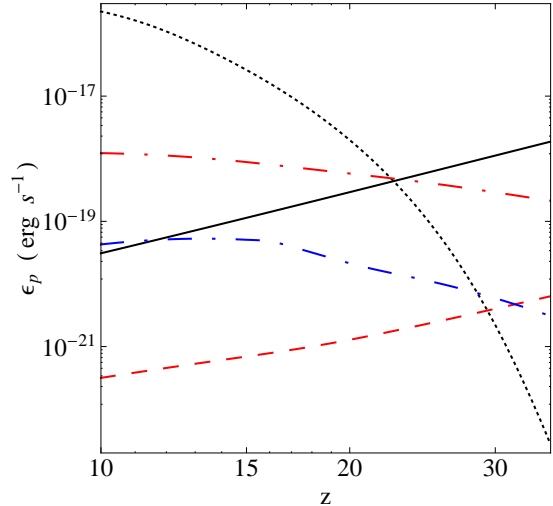}\\
\caption{\small Cooling/heating rates as a function of redshift. We compare here: (i) heating rate $\epsilon_{\rm *}(z)$ from galaxies for the fiducial reionization model (black dotted line); (ii) $\epsilon_{\rm DM}(z)$ for 10 GeV DM annihilations assuming a standard $\langle \sigma v \rangle_{th}$ with and without taking into account the boost from substructures (dot-dashed and dashed red lines respectively); (iii) heating from 200 GeV Wino annihilations including the effect of substructures and assuming $\langle \sigma v \rangle_{th}$ (dot-dashed blue line); (iiii) the adiabatic cooling rate of the expanding gas (solid line). Compton heating is negligible at these redshifts. By coincidence heating from 10 GeV DM annihilations and from galaxies becomes dominant over adiabatic cooling at a similar redshift, $z \sim 20$.}
\label{fig:rates}
\end{figure}

\begin{figure}
\centering
\includegraphics[width=0.49\textwidth]{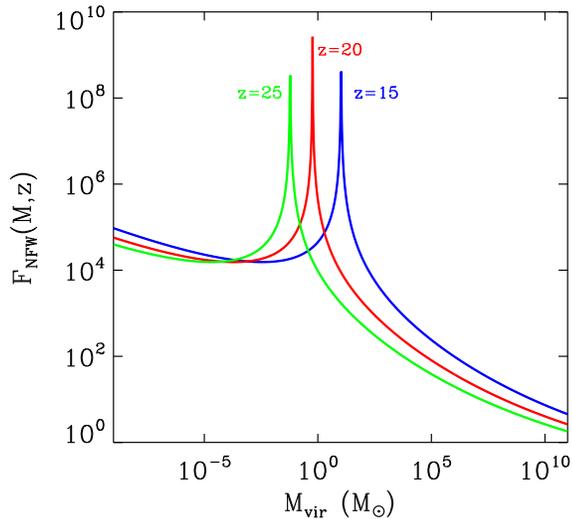}\\
\caption{\small The concentration function adopted to calculate the boost-factor in Eq.~\ref{eq:bz}.}.
\label{fig:concentration}
\end{figure}

It is interesting to notice that purely by coincidence galaxies and DM annihilations start to heat the IGM at about the same redshift in these models even though the processes are entirely different and driven by structures with different mass. As previously mentioned, galaxies with mass $M_{\rm halo} \sim 3 \times 10^{7}$~M$_{\odot}$ ($10^{9}$~M$_{\odot}$) are responsible for most of the astrophysical X-ray heating at $z=20$ for the "fiducial" ("extreme") reionization model. On the other hand the function $F(M,z)$ in Eq. 3 peaks strongly for substructures $M_{\rm sh} \sim 1$~M$_{\odot}$ at the same redshift, as we show in Fig. \ref{fig:concentration}.  The difference in the collapsed fraction above these disparate mass scales is made up by the coefficients in the heating rates.  In Fig.~\ref{fig:rates} we present the heating rates per baryon for 10 GeV DM annihilations ($\epsilon_{\rm DM}$) and for galactic X-ray heating ($\epsilon_{\rm X}$) for the fiducial reionization model, compared with the adiabatic cooling rate (see Eq. 7). Notice that the heating rate from astrophysical sources of X-rays is much steeper than the DM heating, but they become dominant over the adiabatic cooling at about the same redshift. We also plot the heating rate relative to 200 GeV DM annihilations to show that the slope is similar for different DM candidates.

These differences in the slope are easy to understand: the fractional increase of the collapsed fraction in $\gsim 1 M_{\odot}$ halos which drive the DM heating is much slower than the fractional increase in the high-end tail of the mass function (i.e. the halos which host the first galaxies).  This difference is fundamental, and presents an unambiguous way of disentangling heating from astrophysics and heating from DM.

We quantify this further with the dotted red line in the upper panel of Fig.~\ref{fig:dtb}. This curve corresponds to the case of annihilating 10 GeV Binos with $\langle \sigma v \rangle_{th}$, keeping the Ly$\alpha$ pumping due to stellar sources but switching-off astrophysical X-ray heating, which could indeed take place at lower redshifts. The increase of $\delta T_b$ at $z \lsim 22$ ($\nu \sim 60$~MHz) here is only due to heating from DM annihilations: the different slope could be a clear indication of DM heating. 

\begin{figure}
\centering
\includegraphics[width=0.46\textwidth]{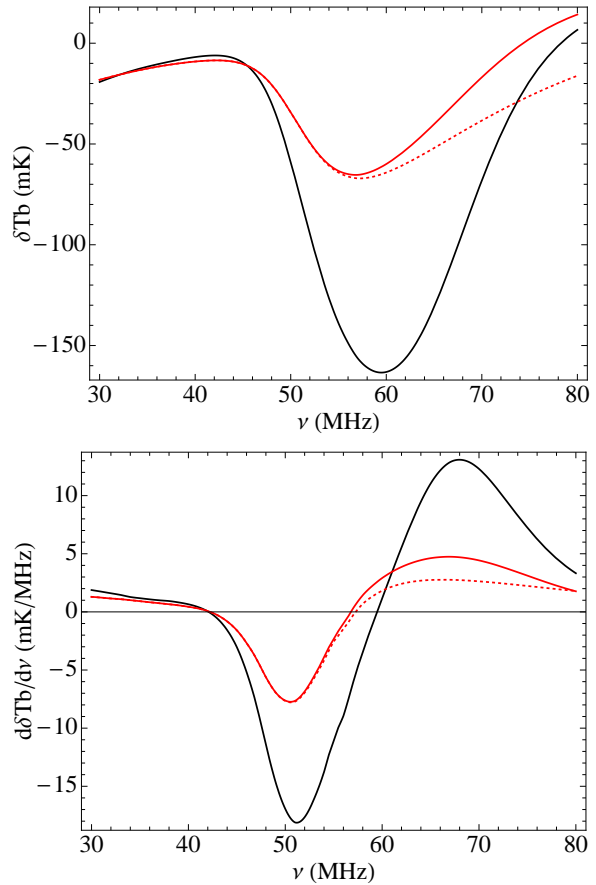}\\
\caption{\small (Top panel) Brightness temperature $\delta T_b$ as a function of frequency $\nu$ assuming the "fiducial" reionization model for three cases: (i) a standard one without DM annihilations (black solid line); (ii) a case including 10 GeV DM annihilations with a standard $\langle \sigma v \rangle_{th}$ and galactic heating and Ly$\alpha$ coupling (red solid line); (iii) a final scenario with 10 GeV DM annihilations with $\langle \sigma v \rangle_{th}$ assuming that galaxies produce Ly$\alpha$ coupling without heating the gas (red dotted line). (Bottom) Gradient $d \delta T_b / d \nu$ of the same quantities. The different slopes relative to heating from DM or galaxies is evident} 
\label{fig:gradient}
\end{figure}

In the bottom panel of Fig.~\ref{fig:gradient} we further investigate the difference in slope between the DM and galactic heating and compare them with the standard scenario in which DM does not annihilate. We study the gradient of the signal with frequency, i.e. $d \delta T_b / d \nu$ in mK$/$MHz, for the 10 GeV Bino, assuming the standard thermal annihilation cross and the fiducial reionization model, and keeping (solid line) or switching off (dotted line) the heating from astrophysical sources. We choose the 10 GeV particle because it is the most promising candidate since it produces a strong signal at $z \sim 20$ with $\langle \sigma v \rangle_{th}$, and we assume the "fiducial" model because in the "extreme" case the X-rays drive the heating and ionization of the IGM and produce with secondary interactions the Ly$\alpha$ flux that couples $T_S$ to $T_K$, therefore it would be incorrect to keep the Ly$\alpha$ pumping and switch off the heating since they are driven by the same sources.

It is evident from the figure that the inclusion of DM heating confines the slope to $d \delta T_b / d \nu \lsim 4$~mK$/$MHz (red curves), whereas without DM heating the slope reaches values of $d \delta T_b / d \nu \sim 14$~mK$/$MHz.  This slope would only increase if rarer (more massive) galaxies drive X-ray heating (such as in the ``extreme'' model).  {\it The detection of such a small peak gradient ($d \delta T_b / d \nu \approx$ few mK$/$MHz) could provide clear evidence of DM annihilation.}

Including the effects of star formation in smaller sized mini-halos would not alter things significantly, since the fractional increase of the collapsed fraction in mini-halos is still much steeper than that relative to the $\gsim 1 M_{\odot}$ sub-halos responsible for the DM heating. In principle, a scenario in which the star formation efficiency within the halos decreases in time (for instance by feedback mechanisms), could reduce the gradient $d \delta T_b / d \nu$. By arbitrarily adjusting the efficiency with time it is possible to perfectly mimic the slower DM heating, however it is unclear how to physically motivate such a rapidly declining star formation efficiency with time. We will explore this aspect in future work.

\subsection{Observational perspectives}

We find that the most promising of our annihilating DM candidates produce a strong deviation in the HI 21~cm line mainly in two redshift ranges: (i) $z \sim 60-200$ (corresponding to $\nu \sim 10-25$~MHz) with a DM signal $\Delta T_{b, DM} \sim 10 - 20$~mK; (ii) $z \lsim 18-30$ ($\nu \sim 45-80$~MHz), where the absorption feature produced by galaxies is strongly suppressed, producing a signal $\Delta T_{b, DM} \sim 20 - 110$~mK depending on the DM particle. 

The first high redshift signature is at $\nu \sim 10-25$~MHz, a frequency too low to be observed by ground based radio interferometers due to the severe distortions produced by terrestrial ionosphere, which at frequencies $\nu \lsim 30$~MHz is known to make any observation virtually impossible. Recently a new generation of space or Moon based radio interferometers able to probe the Dark Ages at very low frequencies has been proposed, such as the Dark Ages Lunar Interferometer (DALI, \citealt{Lazio:2007}), the Lunar Radio Array (LRA, \citealt{Lazio:2009}) and the Dark Ages Radio Explore (DARE, \citealt{Burns:2012}). These instruments, if approved and built, would be located on the dark side of the Moon, overcoming two major hurdles of low frequency radio observations, i.e. radio-frequency interference (RFI) from human-generated radio signals, and ionospheric distortions. They could therefore probe the Dark Ages up to $z \sim 100$.

The second signature at $\nu \sim 45-80$~MHz could be observed by current or planned radio observatories. We focus here on some of the main experiments.

The LOw Frequency ARray (LOFAR), based in the Netherlands, has two observational bands, a low band (30-85 MHz) and a high band (115-230 MHz). The high band array is the one that will be devoted to the EoR experiment, measuring the redshifted HI 21 cm radiation at $6 \lsim z \lsim 11.4$ with a resolution of $\sim$ 3 arcminutes on a field of view of $\sim$ 120 square degrees (\citealt{Zaroubi:2012}). The low band array probes the frequencies at which the DM signal is stronger, however the detection of HI 21~cm signals at such low frequencies will be beyond reach because of a substantial drop in sensitivity (see e.g. \citealt{Zaroubi:2010} and references therein).

The Murchison Widefield Array (MWA), located in the radio-quiet Western Australia Outback, will observe at frequencies from 80 to 300 MHz and therefore misses the frequency bands in which a DM signal could be detected.

The Square Kilometer Array (SKA) which is planned to be completed in 2024 in Australia, New Zealand and South Africa, will probe a frequency range from 70 MHz to 10 GHz. The antenna elements that will make up the "SKA-low" array will cover the frequency range $\nu = 70 - 200$~MHz and will achieve a sensitivity an order of magnitude higher than previous experiments. The design is still preliminary and the telescope is at least a decade away from being fully operational, however SKA could detect a DM signal in the HI 21 cm line, if frequencies down to 70 MHz or lower will be probed.

The main scientific goal of the aforementioned interferometers is however to measure the spatially fluctuating components of the HI 21~cm signal. On the other hand the detection of the angularly averaged all-sky signal could be performed by a single dipole, the main challenge being a precise enough calibration to extract the signal from the backgrounds. This simpler and lower cost alternative is explored by instruments such ad EDGES and the Cosmological Reionization Experiment (CoRE, \citealt{Chippendale:2009}), while more advanced second generation instruments such as the Large-aperture Experiment to Detect the Dark Ages (LEDA) are under construction. A Moon orbiting space observatory such as the Dark Ages Radio Experiment (DARE) seems to be a viable option in the near future, in contrast with the much more complex and expensive task of placing a large interferometer on the dark side of the Moon (DALI, LRA).

The Experiment to Detect the Reionization Step (EDGES) located at the Murchison Radio-astronomy Observatory in Western Australia, measured the radio spectrum between 100 and 200 MHz with all systematic trends in the measurement reduced to below 75 mK, and allowed to exclude a rapid reionization timescale of $\Delta z < 0.06$ at the 95\% confidence level. The EDGES team will attempt in the next few years to reduce the systematics of over one order of magnitude and to push observations up to $z \sim 20$ (see e.g. \citealt{Bowman:2010, McQuinn:2010, Pritchard:2010}).  

One of the most promising instruments for the detection of a DM signal from the Dark Ages is the Large-Aperture Experiment to Detect the Dark Ages (LEDA), a proposed array that will cover a band $\nu \sim 45-90$~MHz and that will feature array-based calibration to improve the accuracy of foreground subtraction from the total-power signal \citep{Greenhill:2012}. The fact that LEDA is optimized for the detection of the all-sky HI 21 cm signal together with its low frequency capabilities make it the ideal ground based instrument to detect the HI 21~cm deviations induced by DM annihilations. 

\section{Summary and conclusions}

\begin{table}
\caption{DM signal for the considered DM models.}
\begin{tabular}{|c|c|c|c|}
\hline
mass [GeV] & $\langle \sigma v \rangle$ & model &Peak $\Delta T_{b,DM}$ [mK]  \\
\hline
200 & $\langle \sigma v \rangle_{th}  $   &   fiducial   &15 mK at $z \sim 23$\\ 
200 & $\langle \sigma v \rangle_{max}$ &   fiducial  &160 mK at $z \sim 23$\\
200 & $\langle \sigma v \rangle_{th}  $    &   extreme  &5 mK at $z \sim 21$\\ 
200 & $\langle \sigma v \rangle_{max}$ &   extreme &45 mK at $z \sim 21$\\
10 & $\langle \sigma v \rangle_{th}$        &   fiducial  &100 mK at $z \sim 23$\\ 
10 & $\langle \sigma v \rangle_{max}$    &   fiducial  &160 mK at $z \sim 23$\\
10 & $\langle \sigma v \rangle_{th}$        &  extreme   &30 mK at $z \sim 21$\\ 
10 & $\langle \sigma v \rangle_{max}$    &  extreme  &45 mK at $z \sim 21$\\
1000 & $\langle \sigma v \rangle_{th}$    &  fiducial  &2 mK at $z \sim 23$\\ 
1000 & $\langle \sigma v \rangle_{max}$&  fiducial  &65 mK at $z \sim 23$\\
1000 & $\langle \sigma v \rangle_{th}$    &   extreme &0.5 mK at $z \sim 21$\\ 
1000 & $\langle \sigma v \rangle_{max}$& extreme  &25 mK at $z \sim 21$\\
\hline
\end{tabular}
\end{table}

We consider three of the most popular annihilating DM candidates as sources of X-ray, ionizing and Ly$\alpha$ radiation: thanks to our detailed Monte Carlo treatment \medea\ we computed their energy deposition into the IGM as a function of redshift and ionized fraction. We were then able to calculate precisely and fully consistently the effects of the energy release from DM annihilations on the IGM thermal history and, ultimately, the imprint on the HI 21~cm line at $20 \lsim z \lsim 200$. For each DM candidate we compute with a modified version of the publicly available code \camb\ the deviations produced by each DM model on the CMB power spectrum. This allows us to select two values for the thermally averaged annihilation cross section, the standard thermal case $\langle \sigma v \rangle_{th} = 3 \times 10^{-26} \mbox{cm}^3 \, \mbox{s}^{-1}$ and the maximum value  $\langle \sigma v \rangle_{max}$ that produces deviations in the CMB power spectrum within $3 \sigma$ of the WMAP 7 results. We combine each DM model with two prescriptions for radiation from astrophysical sources (computed with the public code \cmfast\ ). We investigate how the formation of luminous sources affects the imprint of DM annihilations on the IGM, and study ways to disentangle the DM signal in this realistic scenario.

Finding a strong deviation in the absorption feature in the all-sky averaged $\delta T_b$ at $30 \lsim z \lsim 200$ will be a clear sign of energy release by DM during a cosmic phase in which no collapsed sources of radiation have yet formed. However it is even more interesting to study the joint effects of energy injection into the IGM from galaxies and DM annihilations at $20 \lsim z \lsim 30$, as we find that the Ly$\alpha$ flux from the first luminous sources combined with the early heating from DM annihilations produces strong deviations in the HI 21~cm signal at a redshift range which could be realistically probed by future radio observations in the next few years. Next generation radio facilities such as DALI, LRA or DARE would certainly have ideal characteristics for the detection of a DM signal, but even current or already funded instruments such as MWA, LOFAR and in particular SKA and LEDA would achieve the sensitivity to probe a $\Delta T_{b, DM} \sim 20 - 110$~mK at $\nu \sim 45-80$~MHz, if these frequencies will be observed by their final designs and assuming that the strategies for foreground removal and ionospheric corrections will be successful.

We summarize our results as follows.

\begin{itemize}
\item {\it DM annihilation signal.} \\ (i) Our least massive DM candidate, the 10 GeV Bino like neutralino, produces a large signal in the "fiducial" model, both when assuming the standard thermal cross section $\langle \sigma v \rangle_{th}$ and when taking into account the maximum cross section allowed by CMB data, $\langle \sigma v \rangle_{max}$. In the first case the DM 21~cm signal $\Delta T_{b,DM}$ is of the order of  $5 - 10$ mK on a redshift range $45 \lsim z \lsim 300$ while $\Delta T_{b,DM} \sim 10-100$~mK at $20 \lsim z \lsim 30$, with the peak signal occurring at a frequency $\nu \sim 80$ MHz. Assuming the higher annihilation cross section $\langle \sigma v \rangle_{max}$ increases $\Delta T_{b,DM}$ and totally erases the second absorption feature at $20 \lsim z \lsim 30$, the lack of which represents a very strong - and detectable - DM signature. 
For the "extreme" reionization model we find in general a smaller signal with a peak $\Delta T_{b,DM} \sim 25 \, (50)$~mK at $z \sim 21$ assuming $\langle \sigma v \rangle_{th}$ ($\langle \sigma v \rangle_{max}$).\\
(ii) The annihilating 200 GeV Wino produces very different results depending on the $\langle \sigma v \rangle$ assumption of a thermal or maximal value. In the first case $\Delta T_{b,DM}$ is negligible except for a weak signal $\sim 15$~mK at $z \sim 23$. In the second case instead $\Delta T_{b,DM} \sim 140$~mK at $z \sim 23$ and the second absorption feature driven by Ly$\alpha$ coupling from stars is almost completely erased. For the extreme case we find a peak signal $\Delta T_{b,DM} \sim 5 \, (45)$~mK at $z \sim 21$ for $\langle \sigma v \rangle_{th}$ ($\langle \sigma v \rangle_{max}$). (iii) The results for our most massive candidate, the heavy DM particle of rest mass 1 TeV that pair annihilates into leptons, are the least promising, with a negligible signal under the assumption of $\langle \sigma v \rangle_{th}$, and a $\Delta T_{b,DM} \sim 60 (20)$~mK at $z \sim 23 (100)$. For the extreme case we find a peak signal $\Delta T_{b,DM} \sim 0.5 \, (10)$~mK at $z \sim 21$ for $\langle \sigma v \rangle_{th}$ ($\langle \sigma v \rangle_{max}$).

We summarize our results in Table 2.\\
\item {\it Strategies to isolate the DM signal.} \\ We find that some of our results for a specific DM model working in the framework of the "fiducial" reionization model (e.g. the case of 10 GeV annihilations assuming $\langle \sigma v \rangle_{th}$) are hard to distinguish from the standard signal $\delta T_{b,0}$ assuming the "extreme" reionization history. However when we study the heating rates per baryon for 10 GeV DM annihilations ($\epsilon_{\rm DM}$) and for galactic X-ray heating ($\epsilon_{\rm X}$) we notice that the heating rate from astrophysical sources of X-rays evolves much quicker than the DM heating, since the fractional increase of the collapsed fraction in $\gsim 1 M_{\odot}$ halos which drive the DM heating is much slower than the fractional increase of the halos which host the first galaxies. This crucial difference presents us with a clean way of disentangling heating by DM annihilations from heating by galaxies.
To better quantify this point we study the case of annihilating 10 GeV Binos with $\langle \sigma v \rangle_{th}$, keeping the Ly$\alpha$ pumping due to stellar sources but switching-off astrophysical X-ray heating, which could indeed take place at lower redshifts without breaking any current observational constraints. By coincidence $\delta T_b$ starts increasing at $z \lsim 22$ ($\nu \sim 60$~MHz), same as the case in which source heating is taken into account, however DM heating happens with a different slope. We therefore study the gradient of the signal as a function of frequency, $d \delta T_b / d \nu$, and find that when neglecting DM heating the slope reaches values of $d \delta T_b / d \nu \sim 14$~mK$/$MHz, while in the case in which DM annihilations are taken into account the slope is confined to a considerably smaller $d \delta T_b / d \nu \lsim 4$~mK$/$MHz. Although it is possible in principle to mimic the slower DM heating by arbitrarily reducing the star formation efficiency with time, it is unclear  whether such an "ad-hoc" model could be physically motivated, and we defer this for future analysis.
We can therefore conclude that the detection of a small peak gradient ($d \delta T_b / d \nu \approx$ few mK$/$MHz) will be a strong evidence of DM annihilation, an exciting prospect that could become reality in the next few years thanks to a new generation of radio instruments such as LEDA and DARE, optimized for the detection of the global HI 21~cm signal during the Dark Ages.
\end{itemize}

\section*{Acknowledgments}

We thank J. Chluba and M. Mapelli for their help in using the numerical codes. CE acknowledges a visiting grant from SNS where part of this work has been carried out, and support from the Helmholtz Alliance for Astro-particle Physics funded by the Initiative and Networking Fund of the Helmholtz Association. NY acknowledges the financial support by the Grants-in-Aid for Young Scientists (S) 20674003 by the Japan Society for the Promotion of Science.

\bibliographystyle{mn2e}
\bibliography{vemfy12}

\begin{thebibliography}{88}
\expandafter\ifx\csname natexlab\endcsname\relax\def\natexlab#1{#1}\fi

\bibitem[{{Aalseth} {et~al}\mbox{.}(2011{\natexlab{a}}){Aalseth}, {Barbeau},
  {Bowden}, {Cabrera-Palmer}, {Colaresi}, {Collar}, {Dazeley}, {de Lurgio},
  {Fast}, {Fields}, {Greenberg}, {Hossbach}, {Keillor}, {Kephart}, {Marino},
  {Miley}, {Miller}, {Orrell}, {Radford}, {Reyna}, {Tench}, {van Wechel},
  {Wilkerson}, \& {Yocum}}]{Aalseth:2011a}
{Aalseth} C.~E. {et~al.}, 2011{\natexlab{a}}, Physical Review Letters, 106,
  131301

\bibitem[{{Aalseth} {et~al}\mbox{.}(2011{\natexlab{b}}){Aalseth}, {Barbeau},
  {Colaresi}, {Collar}, {Diaz Leon}, {Fast}, {Fields}, {Hossbach}, {Keillor},
  {Kephart}, {Knecht}, {Marino}, {Miley}, {Miller}, {Orrell}, {Radford},
  {Wilkerson}, \& {Yocum}}]{Aalseth:2011b}
---, 2011{\natexlab{b}}, Physical Review Letters, 107, 141301

\bibitem[{{Abdo} {et~al}\mbox{.}(2010){Abdo}, {Ackermann}, {Ajello}, {Atwood},
  {Baldini}, {Ballet}, {Barbiellini}, {Bastieri}, {Baughman}, {Bechtol},
  {Bellazzini}, {Berenji}, {Blandford}, {Bloom}, {Bonamente}, {Borgland},
  {Bregeon}, {Brez}, {Brigida}, {Bruel}, {Burnett}, {Buson}, {Caliandro},
  {Cameron}, {Caraveo}, {Casandjian}, {Cavazzuti}, {Cecchi}, {{\c C}elik},
  {Charles}, {Chekhtman}, {Cheung}, {Chiang}, {Ciprini}, {Claus},
  {Cohen-Tanugi}, {Cominsky}, {Conrad}, {Cutini}, {Dermer}, {de Angelis}, {de
  Palma}, {Digel}, {di Bernardo}, {E Silva}, {Drell}, {Drlica-Wagner},
  {Dubois}, {Dumora}, {Farnier}, {Favuzzi}, {Fegan}, {Focke}, {Fortin},
  {Frailis}, {Fukazawa}, {Funk}, {Fusco}, {Gaggero}, {Gargano}, {Gasparrini},
  {Gehrels}, {Germani}, {Giebels}, {Giglietto}, {Giommi}, {Giordano},
  {Glanzman}, {Godfrey}, {Grenier}, {Grondin}, {Grove}, {Guillemot}, {Guiriec},
  {Gustafsson}, {Hanabata}, {Harding}, {Hayashida}, {Hughes}, {Itoh},
  {Jackson}, {J{\'o}hannesson}, {Johnson}, {Johnson}, {Johnson}, {Johnson},
  {Kamae}, {Katagiri}, {Kataoka}, {Kawai}, {Kerr}, {Kn{\"o}dlseder}, {Kocian},
  {Kuehn}, {Kuss}, {Lande}, {Latronico}, {Lemoine-Goumard}, {Longo}, {Loparco},
  {Lott}, {Lovellette}, {Lubrano}, {Madejski}, {Makeev}, {Mazziotta},
  {McConville}, {McEnery}, {Meurer}, {Michelson}, {Mitthumsiri}, {Mizuno},
  {Moiseev}, {Monte}, {Monzani}, {Morselli}, {Moskalenko}, {Murgia}, {Nolan},
  {Norris}, {Nuss}, {Ohsugi}, {Omodei}, {Orlando}, {Ormes}, {Paneque},
  {Panetta}, {Parent}, {Pelassa}, {Pepe}, {Pesce-Rollins}, {Piron}, {Porter},
  {Rain{\`o}}, {Rando}, {Razzano}, {Reimer}, {Reimer}, {Reposeur}, {Ritz},
  {Rochester}, {Rodriguez}, {Roth}, {Ryde}, {Sadrozinski}, {Sanchez}, {Sander},
  {Parkinson}, {Scargle}, {Sellerholm}, {Sgr{\`o}}, {Shaw}, {Siskind}, {Smith},
  {Smith}, {Spandre}, {Spinelli}, {Starck}, {Strickman}, {Strong}, {Suson},
  {Tajima}, {Takahashi}, {Takahashi}, {Tanaka}, {Thayer}, {Thayer}, {Thompson},
  {Tibaldo}, {Torres}, {Tosti}, {Tramacere}, {Uchiyama}, {Usher}, {Vasileiou},
  {Vilchez}, {Vitale}, {Waite}, {Wang}, {Winer}, {Wood}, {Ylinen}, {Ziegler},
  \& {Fermi-LAT Collaboration}}]{Abdo:2010a}
{Abdo} A.~A. {et~al.}, 2010, Physical Review Letters, 104, 101101

\bibitem[{{Aprile} {et~al}\mbox{.}(2011){Aprile}, {Arisaka}, {Arneodo},
  {Askin}, {Baudis}, {Behrens}, {Bokeloh}, {Brown}, {Bruch}, {Bruno},
  {Cardoso}, {Chen}, {Choi}, {Cline}, {Duchovni}, {Fattori}, {Ferella}, {Gao},
  {Giboni}, {Gross}, {Kish}, {Lam}, {Lamblin}, {Lang}, {Levy}, {Lim}, {Lin},
  {Lindemann}, {Lindner}, {Lopes}, {Lung}, {Marrod{\'a}n Undagoitia}, {Mei},
  {Melgarejo Fernandez}, {Ni}, {Oberlack}, {Orrigo}, {Pantic}, {Persiani},
  {Plante}, {Ribeiro}, {Santorelli}, {Dos Santos}, {Sartorelli}, {Schumann},
  {Selvi}, {Shagin}, {Simgen}, {Teymourian}, {Thers}, {Vitells}, {Wang},
  {Weber}, \& {Weinheimer}}]{Aprile:2011}
{Aprile} E. {et~al.}, 2011, Physical Review Letters, 107, 131302

\bibitem[{{Barkana} \& {Loeb}(2005)}]{Barkana:2005}
{Barkana} R., {Loeb} A., 2005, \apj, 626, 1

\bibitem[{{Berg} {et~al}\mbox{.}(2009){Berg}, {Edsj{\"o}}, {Gondolo},
  {Lundstr{\"o}m}, \& {Sj{\"o}rs}}]{Berg:2009}
{Berg} M., {Edsj{\"o}} J., {Gondolo} P., {Lundstr{\"o}m} E., {Sj{\"o}rs} S.,
  2009, \jcap, 8, 35

\bibitem[{{Bergstr{\"o}m}, {Edsj{\"o}} \& {Zaharijas}(2009){Bergstr{\"o}m},
  {Edsj{\"o}}, \& {Zaharijas}}]{Bergstrom:2009}
{Bergstr{\"o}m} L., {Edsj{\"o}} J., {Zaharijas} G., 2009, Physical Review
  Letters, 103, 031103

\bibitem[{{Bernabei} {et~al}\mbox{.}(2008){Bernabei}, {Belli}, {Cappella},
  {Cerulli}, {Dai}, {D'Angelo}, {He}, {Incicchitti}, {Kuang}, {Ma},
  {Montecchia}, {Nozzoli}, {Prosperi}, {Sheng}, \& {Ye}}]{Bernabei:2008}
{Bernabei} R. {et~al.}, 2008, European Physical Journal C, 56, 333

\bibitem[{{Bernabei} {et~al}\mbox{.}(2010){Bernabei}, {Belli}, {Cappella},
  {Cerulli}, {Dai}, {D'Angelo}, {He}, {Incicchitti}, {Kuang}, {Ma},
  {Montecchia}, {Nozzoli}, {Prosperi}, {Sheng}, {Wang}, \&
  {Ye}}]{Bernabei:2010}
---, 2010, European Physical Journal C, 67, 39

\bibitem[{{Bernabei} {et~al}\mbox{.}(2004){Bernabei}, {Belli}, {Cappella},
  {Montecchia}, {Nozzoli}, {Incicchitti}, {Prosperi}, {Cerulli}, {Dai},
  {Kuang}, {Ma}, \& {Ye}}]{Bernabei:2004}
---, 2004, arXiv:astro-ph/0405282

\bibitem[{{Bertone} {et~al}\mbox{.}(2012){Bertone}, {Cerde{\~n}o}, {Fornasa},
  {Pieri}, {Ruiz de Austri}, \& {Trotta}}]{Bertone:2012}
{Bertone} G., {Cerde{\~n}o} D.~G., {Fornasa} M., {Pieri} L., {Ruiz de Austri}
  R., {Trotta} R., 2012, \prd, 85, 055014

\bibitem[{{Bowman}, {Morales} \& {Hewitt}(2006){Bowman}, {Morales}, \&
  {Hewitt}}]{Bowman:2006}
{Bowman} J.~D., {Morales} M.~F., {Hewitt} J.~N., 2006, \apj, 638, 20

\bibitem[{{Bowman} \& {Rogers}(2010)}]{Bowman:2010}
{Bowman} J.~D., {Rogers} A.~E.~E., 2010, \nat, 468, 796

\bibitem[{{Bringmann}(2009)}]{Bringmann:2009}
{Bringmann} T., 2009, New Journal of Physics, 11, 105027

\bibitem[{{Burns} {et~al}\mbox{.}(2012){Burns}, {Lazio}, {Bale}, {Bowman},
  {Bradley}, {Carilli}, {Furlanetto}, {Harker}, {Loeb}, \&
  {Pritchard}}]{Burns:2012}
{Burns} J.~O. {et~al.}, 2012, Advances in Space Research, 49, 433

\bibitem[{{CDMS II Collaboration} {et~al}\mbox{.}(2010){CDMS II Collaboration},
  {Ahmed}, {Akerib}, {Arrenberg}, {Bailey}, {Balakishiyeva}, {Baudis}, {Bauer},
  {Brink}, {Bruch}, {Bunker}, {Cabrera}, {Caldwell}, {Cooley}, {Cushman},
  {Daal}, {DeJongh}, {Dragowsky}, {Duong}, {Fallows}, {Figueroa-Feliciano},
  {Filippini}, {Fritts}, {Golwala}, {Grant}, {Hall}, {Hennings-Yeomans},
  {Hertel}, {Holmgren}, {Hsu}, {Huber}, {Kamaev}, {Kiveni}, {Kos}, {Leman},
  {Mahapatra}, {Mandic}, {McCarthy}, {Mirabolfathi}, {Moore}, {Nelson},
  {Ogburn}, {Phipps}, {Pyle}, {Qiu}, {Ramberg}, {Rau}, {Reisetter}, {Saab},
  {Sadoulet}, {Sander}, {Schnee}, {Seitz}, {Serfass}, {Sundqvist}, {Tarka},
  {Wikus}, {Yellin}, {Yoo}, {Young}, \& {Zhang}}]{Ahmed:2010}
{CDMS II Collaboration} {et~al.}, 2010, Science, 327, 1619

\bibitem[{{Chen}, {Takahashi} \& {Yanagida}(2009){Chen}, {Takahashi}, \&
  {Yanagida}}]{Chen:2009}
{Chen} C., {Takahashi} F., {Yanagida} T.~T., 2009, Physics Letters B, 673, 255

\bibitem[{{Chen} \& {Kamionkowski}(2004)}]{Chen:2004}
{Chen} X., {Kamionkowski} M., 2004, \prd, 70, 043502

\bibitem[{{Chippendale}(2009)}]{Chippendale:2009}
{Chippendale} A.~P.~., 2009, PhD thesis, CSIRO Astronomy and Space Science

\bibitem[{{Chluba}(2010)}]{Chluba:2010}
{Chluba} J., 2010, \mnras, 402, 1195

\bibitem[{{Ciafaloni} {et~al}\mbox{.}(2011){Ciafaloni}, {Comelli}, {Riotto},
  {Sala}, {Strumia}, \& {Urbano}}]{Ciafaloni:2011}
{Ciafaloni} P., {Comelli} D., {Riotto} A., {Sala} F., {Strumia} A., {Urbano}
  A., 2011, \jcap, 3, 19

\bibitem[{{Cirelli}, {Franceschini} \& {Strumia}(2008){Cirelli},
  {Franceschini}, \& {Strumia}}]{Cirelli:2008}
{Cirelli} M., {Franceschini} R., {Strumia} A., 2008, Nuclear Physics B, 800,
  204

\bibitem[{{Cirelli}, {Iocco} \& {Panci}(2009){Cirelli}, {Iocco}, \&
  {Panci}}]{Cirelli:2009}
{Cirelli} M., {Iocco} F., {Panci} P., 2009, \jcap, 10, 9

\bibitem[{{Cirelli} {et~al}\mbox{.}(2009){Cirelli}, {Kadastik}, {Raidal}, \&
  {Strumia}}]{Cirelli:2009b}
{Cirelli} M., {Kadastik} M., {Raidal} M., {Strumia} A., 2009, Nuclear Physics
  B, 813, 1

\bibitem[{{di Bernardo} {et~al}\mbox{.}(2011){di Bernardo}, {Evoli}, {Gaggero},
  {Grasso}, {Maccione}, \& {Mazziotta}}]{diBernardo:2011}
{di Bernardo} G., {Evoli} C., {Gaggero} D., {Grasso} D., {Maccione} L.,
  {Mazziotta} M.~N., 2011, Astroparticle Physics, 34, 528

\bibitem[{{Evoli} {et~al}\mbox{.}(2012{\natexlab{a}}){Evoli}, {Cholis},
  {Grasso}, {Maccione}, \& {Ullio}}]{Evoli:2011}
{Evoli} C., {Cholis} I., {Grasso} D., {Maccione} L., {Ullio} P.,
  2012{\natexlab{a}}, \prd, 85, 123511

\bibitem[{{Evoli} {et~al}\mbox{.}(2012{\natexlab{b}}){Evoli}, {Vald{\'e}s},
  {Ferrara}, \& {Yoshida}}]{Evoli:2012}
{Evoli} C., {Vald{\'e}s} M., {Ferrara} A., {Yoshida} N., 2012{\natexlab{b}},
  \mnras, 422, 420

\bibitem[{{Field}(1959)}]{Field:1959}
{Field} G.~B., 1959, \apj, 129, 536

\bibitem[{{Finkbeiner} {et~al}\mbox{.}(2012){Finkbeiner}, {Galli}, {Lin}, \&
  {Slatyer}}]{Finkbeiner:2012}
{Finkbeiner} D.~P., {Galli} S., {Lin} T., {Slatyer} T.~R., 2012, \prd, 85,
  043522

\bibitem[{{Furlanetto}(2006)}]{Furlanetto:2006a}
{Furlanetto} S.~R., 2006, \mnras, 371, 867

\bibitem[{{Furlanetto}, {Oh} \& {Briggs}(2006){Furlanetto}, {Oh}, \&
  {Briggs}}]{Furlanetto:2006r}
{Furlanetto} S.~R., {Oh} S.~P., {Briggs} F.~H., 2006, \physrep, 433, 181

\bibitem[{{Furlanetto}, {Oh} \& {Pierpaoli}(2006){Furlanetto}, {Oh}, \&
  {Pierpaoli}}]{Furlanetto:2006}
{Furlanetto} S.~R., {Oh} S.~P., {Pierpaoli} E., 2006, \prd, 74, 103502

\bibitem[{{Galli} {et~al}\mbox{.}(2009){Galli}, {Iocco}, {Bertone}, \&
  {Melchiorri}}]{Galli:2009}
{Galli} S., {Iocco} F., {Bertone} G., {Melchiorri} A., 2009, \prd, 80, 023505

\bibitem[{{Galli} {et~al}\mbox{.}(2011){Galli}, {Iocco}, {Bertone}, \&
  {Melchiorri}}]{Galli:2011}
---, 2011, \prd, 84, 027302

\bibitem[{{Gilfanov}, {Grimm} \& {Sunyaev}(2004){Gilfanov}, {Grimm}, \&
  {Sunyaev}}]{Gilfanov:2004}
{Gilfanov} M., {Grimm} H.-J., {Sunyaev} R., 2004, \mnras, 347, L57

\bibitem[{{Grajek} {et~al}\mbox{.}(2009){Grajek}, {Kane}, {Phalen}, {Pierce},
  \& {Watson}}]{Grajek:2009}
{Grajek} P., {Kane} G.~L., {Phalen} D.~J., {Pierce} A., {Watson} S., 2009,
  \prd, 79, 043506

\bibitem[{{Greenhill} \& {Bernardi}(2012)}]{Greenhill:2012}
{Greenhill} L.~J., {Bernardi} G., 2012, arXiv:1201.1700

\bibitem[{{Hickox} \& {Markevitch}(2007)}]{HM07}
{Hickox} R.~C., {Markevitch} M., 2007, \apjl, 661, L117

\bibitem[{{Hirata}(2006)}]{Hirata:2006}
{Hirata} C.~M., 2006, \mnras, 367, 259

\bibitem[{{Hooper} \& {Tait}(2009)}]{Hooper:2009}
{Hooper} D., {Tait} T.~M.~P., 2009, \prd, 80, 055028

\bibitem[{{Jenkins} {et~al}\mbox{.}(2001){Jenkins}, {Frenk}, {White},
  {Colberg}, {Cole}, {Evrard}, {Couchman}, \& {Yoshida}}]{Jenkins:2001}
{Jenkins} A., {Frenk} C.~S., {White} S.~D.~M., {Colberg} J.~M., {Cole} S.,
  {Evrard} A.~E., {Couchman} H.~M.~P., {Yoshida} N., 2001, \mnras, 321, 372

\bibitem[{{Kane}, {Lu} \& {Watson}(2009){Kane}, {Lu}, \& {Watson}}]{Kane:2009}
{Kane} G., {Lu} R., {Watson} S., 2009, Physics Letters B, 681, 151

\bibitem[{{Kassim} {et~al}\mbox{.}(2004){Kassim}, {Lazio}, {Ray}, {Crane},
  {Hicks}, {Stewart}, {Cohen}, \& {Lane}}]{Kassim:2004}
{Kassim} N.~E., {Lazio} T.~J.~W., {Ray} P.~S., {Crane} P.~C., {Hicks} B.~C.,
  {Stewart} K.~P., {Cohen} A.~S., {Lane} W.~M., 2004, \planss, 52, 1343

\bibitem[{{Komatsu} {et~al}\mbox{.}(2009){Komatsu}, {Dunkley}, {Nolta},
  {Bennett}, {Gold}, {Hinshaw}, {Jarosik}, {Larson}, {Limon}, {Page},
  {Spergel}, {Halpern}, {Hill}, {Kogut}, {Meyer}, {Tucker}, {Weiland},
  {Wollack}, \& {Wright}}]{Komatsu:2009}
{Komatsu} E. {et~al.}, 2009, \apjs, 180, 330

\bibitem[{{Komatsu} {et~al}\mbox{.}(2011){Komatsu}, {Smith}, {Dunkley},
  {Bennett}, {Gold}, {Hinshaw}, {Jarosik}, {Larson}, {Nolta}, {Page},
  {Spergel}, {Halpern}, {Hill}, {Kogut}, {Limon}, {Meyer}, {Odegard}, {Tucker},
  {Weiland}, {Wollack}, \& {Wright}}]{Komatsu:2011}
---, 2011, \apjs, 192, 18

\bibitem[{Labbe {et~al}\mbox{.}(2010)Labbe {et~al.}}]{Labbe:2010}
Labbe I., {et~al.}, 2010, \apjl, 708, L26

\bibitem[{{Lazio} {et~al}\mbox{.}(2009){Lazio}, {Carilli}, {Hewitt},
  {Furlanetto}, \& {Burns}}]{Lazio:2009}
{Lazio} J., {Carilli} C., {Hewitt} J., {Furlanetto} S., {Burns} J., 2009, in
  Society of Photo-Optical Instrumentation Engineers (SPIE) Conference Series,
  Vol. 7436, Society of Photo-Optical Instrumentation Engineers (SPIE)
  Conference Series

\bibitem[{{Lazio} {et~al}\mbox{.}(2007){Lazio}, {Kasper}, {Jones}, {Burns},
  {Furlanetto}, {Weiler}, {MacDowall}, {Demaio}, {Bale}, {Ellingson},
  {Greenhill}, \& {Taylor}}]{Lazio:2007}
{Lazio} T.~J.~W. {et~al.}, 2007, in Bulletin of the American Astronomical
  Society, Vol.~39, American Astronomical Society Meeting Abstracts

\bibitem[{{Lewis} \& {Challinor}(2011)}]{Lewis:2011}
{Lewis} A., {Challinor} A., 2011, in Astrophysics Source Code Library, record
  ascl:1102.026, p. 2026

\bibitem[{{Linden}, {Profumo} \& {Anderson}(2010){Linden}, {Profumo}, \&
  {Anderson}}]{Linden:2010}
{Linden} T., {Profumo} S., {Anderson} B., 2010, \prd, 82, 063529

\bibitem[{{Liu} {et~al}\mbox{.}(2010){Liu}, {Yuan}, {Bi}, {Li}, \&
  {Zhang}}]{Liu:2009}
{Liu} J., {Yuan} Q., {Bi} X., {Li} H., {Zhang} X., 2010, \prd, 81, 023516

\bibitem[{{Lutovinov} {et~al}\mbox{.}(2005){Lutovinov}, {Revnivtsev},
  {Gilfanov}, {Shtykovskiy}, {Molkov}, \& {Sunyaev}}]{Lutovinov:2005}
{Lutovinov} A., {Revnivtsev} M., {Gilfanov} M., {Shtykovskiy} P., {Molkov} S.,
  {Sunyaev} R., 2005, \aap, 444, 821

\bibitem[{{Madau}, {Meiksin} \& {Rees}(1997){Madau}, {Meiksin}, \&
  {Rees}}]{Madau:1997}
{Madau} P., {Meiksin} A., {Rees} M.~J., 1997, \apj, 475, 429

\bibitem[{{Madau} {et~al}\mbox{.}(2004){Madau}, {Rees}, {Volonteri}, {Haardt},
  \& {Oh}}]{Madau:2004}
{Madau} P., {Rees} M.~J., {Volonteri} M., {Haardt} F., {Oh} S.~P., 2004, \apj,
  604, 484

\bibitem[{{Mapelli}, {Ferrara} \& {Pierpaoli}(2006){Mapelli}, {Ferrara}, \&
  {Pierpaoli}}]{Mapelli:2006}
{Mapelli} M., {Ferrara} A., {Pierpaoli} E., 2006, \mnras, 369, 1719

\bibitem[{{McQuinn}(2010)}]{McQuinn:2010}
{McQuinn} M., 2010, in Astronomical Society of the Pacific Conference Series,
  Vol. 432, New Horizons in Astronomy: Frank N. Bash Symposium 2009, {Stanford}
  L.~M., {Green} J.~D., {Hao} L., {Mao} Y., eds., p.~65

\bibitem[{{McQuinn}(2012)}]{McQuinn:2012}
---, 2012, arXiv:1206.1335

\bibitem[{{Mesinger} \& {Dijkstra}(2008)}]{Mesinger:2008}
{Mesinger} A., {Dijkstra} M., 2008, \mnras, 390, 1071

\bibitem[{{Mesinger}, {Ferrara} \& {Spiegel}(2012){Mesinger}, {Ferrara}, \&
  {Spiegel}}]{Mesinger:2012a}
{Mesinger} A., {Ferrara} A., {Spiegel} D.~S., 2012, arXiv:1210.7319

\bibitem[{{Mesinger} \& {Furlanetto}(2007)}]{Mesinger:2007}
{Mesinger} A., {Furlanetto} S., 2007, \apj, 669, 663

\bibitem[{{Mesinger}, {Furlanetto} \& {Cen}(2011){Mesinger}, {Furlanetto}, \&
  {Cen}}]{Mesinger:2011}
{Mesinger} A., {Furlanetto} S., {Cen} R., 2011, \mnras, 411, 955

\bibitem[{{Mesinger}, {McQuinn} \& {Spergel}(2012){Mesinger}, {McQuinn}, \&
  {Spergel}}]{Mesinger:2012}
{Mesinger} A., {McQuinn} M., {Spergel} D.~N., 2012, \mnras, 422, 1403

\bibitem[{{Mineo}, {Gilfanov} \& {Sunyaev}(2012){Mineo}, {Gilfanov}, \&
  {Sunyaev}}]{Mineo:2012}
{Mineo} S., {Gilfanov} M., {Sunyaev} R., 2012, \mnras, 419, 2095

\bibitem[{{Mirabel} {et~al}\mbox{.}(2011){Mirabel}, {Dijkstra}, {Laurent},
  {Loeb}, \& {Pritchard}}]{Mirabel:2011}
{Mirabel} I.~F., {Dijkstra} M., {Laurent} P., {Loeb} A., {Pritchard} J.~R.,
  2011, \aap, 528, A149

\bibitem[{{Natarajan} \& {Schwarz}(2009)}]{Natarajan:2009}
{Natarajan} A., {Schwarz} D.~J., 2009, \prd, 80, 043529

\bibitem[{{Okamoto}, {Gao} \& {Theuns}(2008){Okamoto}, {Gao}, \&
  {Theuns}}]{Okamoto:2008}
{Okamoto} T., {Gao} L., {Theuns} T., 2008, \mnras, 390, 920

\bibitem[{{Peterson}, {Pen} \& {Wu}(2005){Peterson}, {Pen}, \&
  {Wu}}]{Peterson:2005}
{Peterson} J.~B., {Pen} U., {Wu} X., 2005, in Astronomical Society of the
  Pacific Conference Series, Vol. 345, From Clark Lake to the Long Wavelength
  Array: Bill Erickson's Radio Science, {Kassim} N., {Perez} M., {Junor} W.,
  {Henning} P., eds., p. 441

\bibitem[{{Press} \& {Schechter}(1974)}]{Press:1974}
{Press} W.~H., {Schechter} P., 1974, \apj, 187, 425

\bibitem[{{Pritchard} \& {Furlanetto}(2007)}]{Pritchard:2007}
{Pritchard} J.~R., {Furlanetto} S.~R., 2007, \mnras, 376, 1680

\bibitem[{{Pritchard} \& {Loeb}(2008)}]{Pritchard:2008}
{Pritchard} J.~R., {Loeb} A., 2008, \prd, 78, 103511

\bibitem[{{Pritchard} \& {Loeb}(2010)}]{Pritchard:2010}
---, 2010, \prd, 82, 023006

\bibitem[{{Profumo}(2012)}]{Profumo:2008}
{Profumo} S., 2012, Central European Journal of Physics, 10, 1

\bibitem[{{Reichardt} {et~al}\mbox{.}(2011){Reichardt}, {Shaw}, {Zahn}, {Aird},
  {Benson}, {Bleem}, {Carlstrom}, {Chang}, {Cho}, {Crawford}, {Crites}, {de
  Haan}, {Dobbs}, {Dudley}, {George}, {Halverson}, {Holder}, {Holzapfel},
  {Hoover}, {Hou}, {Hrubes}, {Joy}, {Keisler}, {Knox}, {Lee}, {Leitch},
  {Lueker}, {Luong-Van}, {McMahon}, {Mehl}, {Meyer}, {Millea}, {Mohr},
  {Montroy}, {Natoli}, {Padin}, {Plagge}, {Pryke}, {Ruhl}, {Schaffer},
  {Shirokoff}, {Spieler}, {Staniszewski}, {Stark}, {Story}, {van Engelen},
  {Vanderlinde}, {Vieira}, \& {Williamson}}]{Reichardt:2011}
{Reichardt} C.~L. {et~al.}, 2011, arXiv:1111.0932

\bibitem[{{Ricotti} \& {Ostriker}(2004)}]{RO:2004}
{Ricotti} M., {Ostriker} J.~P., 2004, \mnras, 352, 547

\bibitem[{{Shchekinov} \& {Vasiliev}(2007)}]{Shchekinov:2007}
{Shchekinov} Y.~A., {Vasiliev} E.~O., 2007, \mnras, 379, 1003

\bibitem[{{Sheth} \& {Tormen}(1999)}]{Sheth:1999}
{Sheth} R.~K., {Tormen} G., 1999, \mnras, 308, 119

\bibitem[{{Shull} {et~al}\mbox{.}(2012){Shull}, {Harness}, {Trenti}, \&
  {Smith}}]{Shull:2012}
{Shull} J.~M., {Harness} A., {Trenti} M., {Smith} B.~D., 2012, \apj, 747, 100

\bibitem[{{Slatyer}, {Padmanabhan} \& {Finkbeiner}(2009){Slatyer},
  {Padmanabhan}, \& {Finkbeiner}}]{Slatyer:2009}
{Slatyer} T.~R., {Padmanabhan} N., {Finkbeiner} D.~P., 2009, \prd, 80, 043526

\bibitem[{{Springel} \& {Hernquist}(2003)}]{Springel:2003}
{Springel} V., {Hernquist} L., 2003, \mnras, 339, 312

\bibitem[{{The Planck Collaboration}(2006)}]{Planck:2006}
{The Planck Collaboration}, 2006, arXiv:astro-ph/0604069

\bibitem[{{Vald{\'e}s} \& {Ferrara}(2008)}]{Valdes:2008}
{Vald{\'e}s} M., {Ferrara} A., 2008, \mnras, 387, L8

\bibitem[{{Vald{\'e}s} {et~al}\mbox{.}(2007){Vald{\'e}s}, {Ferrara}, {Mapelli},
  \& {Ripamonti}}]{Valdes:2007}
{Vald{\'e}s} M., {Ferrara} A., {Mapelli} M., {Ripamonti} E., 2007, \mnras, 377,
  245

\bibitem[{{van den Aarssen}, {Bringmann} \& {Goedecke}(2012){van den Aarssen},
  {Bringmann}, \& {Goedecke}}]{Aarssen:2012}
{van den Aarssen} L.~G., {Bringmann} T., {Goedecke} Y.~C., 2012,
  arXiv:1202.5456

\bibitem[{{Wouthuysen}(1952)}]{Wouthuysen:1952}
{Wouthuysen} S.~A., 1952, \aj, 57, 31

\bibitem[{{Wyithe}, {Loeb} \& {Barnes}(2005){Wyithe}, {Loeb}, \&
  {Barnes}}]{Wyithe:2005}
{Wyithe} J.~S.~B., {Loeb} A., {Barnes} D.~G., 2005, \apj, 634, 715

\bibitem[{{Zahn} {et~al}\mbox{.}(2011){Zahn}, {Mesinger}, {McQuinn}, {Trac},
  {Cen}, \& {Hernquist}}]{Zahn:2011}
{Zahn} O., {Mesinger} A., {McQuinn} M., {Trac} H., {Cen} R., {Hernquist} L.~E.,
  2011, \mnras, 414, 727

\bibitem[{{Zaroubi}(2010)}]{Zaroubi:2010}
{Zaroubi} S., 2010, arXiv:1002.2667

\bibitem[{{Zaroubi}(2012)}]{Zaroubi:2012}
---, 2012, arXiv:1206.0267

\end{thebibliography}

\end{document}